\documentclass[a4paper,twocolumn,11pt,accepted=2022-06-02]{quantumarticle}
\pdfoutput=1
\usepackage[utf8]{inputenc}
\usepackage[english]{babel}
\usepackage[T1]{fontenc}
\usepackage{amsmath}
\usepackage{hyperref}
\usepackage[numbers,sort&compress]{natbib}
\usepackage{tikz}
\usepackage{lipsum}
\usepackage{braket}
\usepackage{verbatim}
\usepackage{hyperref}
\usepackage{cleveref}
\usepackage{multirow}
\usepackage{bbm}

\DeclareMathOperator{\tr}{tr}

\begin{document}

\author{Pieter W. Claeys}
\email{claeys@pks.mpg.de}
\affiliation{Max Planck Institute for the Physics of Complex Systems, 01187 Dresden, Germany}
\affiliation{TCM Group, Cavendish Laboratory, University of Cambridge, Cambridge CB3 0HE, UK}
\orcid{0000-0001-7150-8459}

\author{Austen Lamacraft}
\affiliation{TCM Group, Cavendish Laboratory, University of Cambridge, Cambridge CB3 0HE, UK}
\orcid{0000-0002-0707-0488}

\title{Emergent quantum state designs and biunitarity in dual-unitary circuit dynamics}

\maketitle

\begin{abstract}
Recent works have investigated the emergence of a new kind of random matrix behaviour in unitary dynamics following a quantum quench. Starting from a time-evolved state, an ensemble of pure states supported on a small subsystem can be generated by performing projective measurements on the remainder of the system, leading to a \emph{projected ensemble}. In chaotic quantum systems it was conjectured that such projected ensembles become indistinguishable from the uniform Haar-random ensemble and lead to a \emph{quantum state design}. Exact results were recently presented by Ho and Choi [\href{https://doi.org/10.1103/PhysRevLett.128.060601}{Phys. Rev. Lett. 128, 060601 (2022)}] for the kicked Ising model at the self-dual point. We provide an alternative construction that can be extended to general chaotic dual-unitary circuits with solvable initial states and measurements, highlighting the role of the underlying dual-unitarity and further showing how dual-unitary circuit models exhibit both exact solvability and random matrix behaviour. Building on results from biunitary connections, we show how complex Hadamard matrices and unitary error bases both lead to solvable measurement schemes.
\end{abstract}

\section{Introduction}

In many-body quantum dynamics the notions of quantum chaos and thermalization are intimately connected to random matrix theory (RMT) \cite{d2016quantum}. Generic closed chaotic quantum systems are expected to display many universal behaviours characteristic of ensembles of random matrices, and the presence of random matrix level statistics is currently seen as one of the defining characteristics of quantum chaos \cite{d2016quantum,stockmann_quantum_1999,haake_quantum_2010}.
While there is a large body of numerical evidence for this connection, chaotic behaviour typically precludes any exact treatment of the dynamics.

In recent years, \emph{dual-unitary circuits} have emerged as a class of unitary models for which the dynamics of correlations remains tractable. These models are characterized by an underlying space-time duality and can be either integrable or chaotic. 
Interest in dual-unitary models originated with various studies of the kicked Ising model (KIM) \cite{akila_particle-time_2016}, which is amenable to calculation of the spectral form factor and entanglement entropies at particular values of the coupling constants \cite{bertini_exact_2018,bertini_entanglement_2019}. Ref.~\cite{gopalakrishnan_unitary_2019} recast this model in terms of (dual-)unitary circuits, and Ref.~\cite{bertini_exact_2019} showed how dynamical correlation functions could be exactly calculated for dual-unitary circuits and classified all such circuits for a local two-dimensional Hilbert space (qubits). No complete characterization of dual-unitary gates exists for arbitrary Hilbert spaces, but various constructions have been proposed leading to rich dynamical behaviours \cite{rather_creating_2020,gutkin_exact_2020,aravinda_dual-unitary_2021,claeys_ergodic_2021,prosen_many-body_2021,singh_diagonal_2021,borsi_remarks_2022}. 
Subsequent studies led to exact results on scrambling \cite{claeys_maximum_2020,bertini_scrambling_2020}, computational power \cite{suzuki_computational_2022}, solvable initial matrix product states \cite{piroli_exact_2020}, entanglement measures \cite{jonnadula_entanglement_2020} and entanglement barriers \cite{reid_entanglement_2021}, the effect of perturbations \cite{kos_correlations_2021}, the evolution of temporal entanglement \cite{lerose_influence_2021,giudice_temporal_2021,lerose_overcoming_2022}, and measurement-induced phase transitions \cite{zabalo_operator_2022}. With the advent of quantum simulators, recent experiments have directly probed dynamical properties in dual-unitary models, including correlations \cite{chertkov_holographic_2021} and scrambling \cite{mi_information_2021}.

Various spectral properties have also been analytically calculated in these models, confirming predictions from RMT and indicating quantum chaotic behaviour.
Bertini \emph{et al.} provided the first proof of the emergence of the RMT spectral form factor for the KIM \cite{bertini_exact_2018}, which was subsequently generalized to a broader class of dual-unitary dynamics \cite{bertini_random_2021}. 
In the presence of noisy driving, the unitary evolution operator was argued to reduce to a random matrix in the many-body Hilbert space through the calculation of a generalized spectral form factor \cite{kos_chaos_2021}.
Dual-unitary circuits have also allowed for the calculation of the asymptotic form of the spectral function, confirming predictions from the eigenstate thermalization hypothesis \cite{fritzsch_eigenstate_2021}.

A different kind of emergent random matrix behaviour was recently proposed by Cotler \emph{et al.} \cite{cotler_emergent_2021}, going beyond the standard paradigm of thermalization. Starting from a single many-body wave function, an ensemble of pure states on a subsystem can be generated by performing local measurements over part of the system. Partitioning a full lattice in a subsystem A and a bath B, performing local measurements on B leads to both a probability for the measurement outcome and a pure state supported on A, with the resulting ensemble known as a \emph{projected ensemble}. In Ref.~\cite{cotler_emergent_2021} it was argued that for generic many-body quantum states such ensembles give rise to emergent \emph{quantum state $k$-designs}, i.e. the first $k$-moments of the ensemble coincide with those of the uniform distribution over the full Hilbert space. In a quantum quench, for longer time evolution the time-evolved states were numerically shown to form higher (approximate) $k$-designs. This emergence of $k$-designs was simultaneously experimentally observed in a Rydberg quantum simulator \cite{choi_emergent_2021}.

In Ref.~\cite{ho_exact_2021} Ho and Choi presented the first analytical proof for emergent quantum state designs by considering the KIM at the self-dual point. In this setup, a stronger emergence was shown: in the thermodynamic limit of an infinite bath size all moments of the projected ensemble are indistinguishable from those of the uniform Haar distribution after a finite time evolution. In other words, the distribution of quantum states due to measurements of the bath state is indistinguishable from the purely uniform Haar-random distribution and the projected ensemble is said to form a \emph{quantum state design} \cite{gross_evenly_2007,ambainis_quantum_2007,roberts_chaos_2017}. The underlying proof relies on identifying a unitary transfer matrix generating evolution along the space direction depending on the local measurement outcome, and showing that the different measurement outcomes produce a universal gate set.

Wilming and Roth subsequently showed the emergence of approximate $k$-designs in every system where a small subsystem is close to maximally mixed \cite{wilming_high-temperature_2022}. In this way thermalization of the reduced density matrix to infinite temperature implies the emergence of quantum state designs provided the measurement basis is chosen uniformly at random from the Haar measurement. The emergence of quantum state designs can be seen as the measurement becoming more and more `random' w.r.t. the time-evolved state at longer times. 

We present an alternative construction for emergent quantum state designs in dual-unitary circuit dynamics, more in line with the recent literature on dual-unitarity. This approach makes explicit in what way the presented construction depends on the choice of unitary gates, initial state, and measurement basis. We use the recent connection between dual-unitarity and \emph{biunitarity} as pointed out in Ref.~\cite{borsi_remarks_2022} to extend these results in different directions. Biunitarity encompasses dual-unitary gates as well as complex Hadamard matrices, unitary error bases, and quantum Latin squares, endowing all such constructions with properties that go beyond simple unitarity  \cite{reutter_biunitary_2019}. We provide two examples of solvable measurement schemes based on biunitary constructions: 1) keeping the measurements in the single-site basis and constructing dual-unitary gates out of complex Hadamard matrices and 2) considering measurements corresponding to a unitary error basis on neighbouring sites (e.g. Bell pair states) for generic `maximally chaotic' dual-unitary gates. In both cases all moments of the projected ensemble are  argued to reproduce the moments of the uniform distribution after a time proportional to the subsystem size, and generate a quantum state design for these `atypical' measurement bases. 

In the remainder of this introduction we will give the precise definitions of quantum state designs (Sec.~\ref{subsec:definitions_ensembles}) and the unitary circuit dynamics that are the focus of this work (Sec.~\ref{subsec:unitary_circuits}). The specific case of dual-unitary circuit dynamics is discussed in Sec.~\ref{sec:design_dualunitary}, outlining the transfer matrix approach for the calculation of the moments of the projected ensemble. Solvable measurement schemes from complex Hadamard matrices and unitary error bases are introduced in Secs.~\ref{subsec:Hadamard} and \ref{subsec:solvable_meas} respectively. The resulting state designs are illustrated numerically in Sec.~\ref{sec:numerics} and contrasted with results for generic unitary and dual-unitary dynamics. Secs.~\ref{sec:discussion} and ~\ref{sec:conclusions} present a discussion and conclusions.

\subsection{Quantum state designs and projected ensembles}
\label{subsec:definitions_ensembles}
Following Refs.~\cite{cotler_emergent_2021,choi_emergent_2021,ho_exact_2021} the projected ensemble is defined from a single quantum state $\ket{\psi}$ for a lattice consisting of $N$ qubits or spin-1/2 subsystems. The lattice can be partitioned in a system $A$ and a `bath' $B$ consisting of $N_A$ and $N_B$ sites respectively. 
A projective measurement of the bath state in e.g. the computational basis will return an outcome $z_B = (z_{1},z_2,\dots,z_{N_B}) \in \{0,1\}^{N_B}$ with probability $p(z_B)$ and an associated pure state $\ket{\psi(z_B)}$ on $A$ given by
\begin{align}
p(z_B) &= \bra{\psi} \left(\mathbbm{1}_A \otimes |{z_B}\rangle\langle z_B | \right) \ket{\psi} \\
\ket{\psi(z_B)} &= \left(\mathbbm{1}_A \otimes \bra{z_B}\right)\ket{\psi}/\sqrt{p(z_B)}
\end{align}
The set of all probabilities and corresponding states forms the projected ensemble $\mathcal{E}=\{p(z_B),\ket{\psi(z_B)}\}$.
The reduced density matrix for $A$ can be obtained as the first moment of this ensemble, i.e.
\begin{align}
\rho_A &= \mathrm{Tr}_B\big(\ket{\psi}\bra{\psi}\big) \nonumber\\
&= \sum_{z_B} p(z_B) \ket{\psi(z_B)}\bra{\psi(z_B)},
\end{align}
and higher moments are defined as
\begin{align}
\rho_{\mathcal{E}}^{(k)} = \sum_{z_B}p(z_B) \left(\ket{\psi(z_B)}\bra{\psi(z_B)}\right)^{\otimes k}\,.
\end{align}
Note that thermalization is observed purely in the reduced density matrix and hence the first moment of this ensemble, while quantities involving multiple copies of the system will depend on the higher moments.

In order for this ensemble to be a quantum state design, all $k$-th moments should agree with those for the ensemble of uniformly (Haar-random) distributed states on $A$, as given by (see e.g. \cite{roberts_chaos_2017})
\begin{align}
\rho_{\textrm{Haar}}^{(k)} &= \int d\phi \left(\ket{\phi}\bra{\phi}\right)^{\otimes k} \nonumber\\
&= \frac{\sum_{\pi \in S_k}P(\pi)}{d(d+1)\dots(d+k-1)}\,.
\end{align}
The states $\ket{\phi}$ are Haar random states in a $d=2^{N_A}$ dimensional Hilbert space, and $P(\pi)$ is an operator acting on $k$ copies of this Hilbert space, permuting these copies according to a permutation $\pi \in S_k$ of $k$ elements, i.e.
\begin{align}\label{eq:def_P_pi}
P(\pi)\ket{i_1, i_2,\dots,i_k} = \ket{i_{\pi(1)},i_{\pi(2)}, \dots, i_{\pi(k)}}.
\end{align}

\subsection{Unitary circuit dynamics}
\label{subsec:unitary_circuits}
In the remainder of this work, we will consider projected ensembles from states that are obtained using unitary circuit dynamics. Such unitary circuits originated in quantum computation and have gained popularity as a minimal model for local many-body dynamics \cite{chandran_semiclassical_2015,nahum_quantum_2017,khemani_operator_2018,von_keyserlingk_operator_2018,nahum_operator_2018,chan_solution_2018,rakovszky_diffusive_2018,rakovszky_sub-ballistic_2019,zhou_entanglement_2020,garratt_local_2021,bensa_fastest_2021,lerose_influence_2021}. The unitary evolution operator is represented as a unitary circuit consisting of two-site operators, where each gate $U$ and its hermitian conjugate can be graphically represented as
\begin{align}\label{eq:def_U_Udag}
U_{ab,cd} = \vcenter{\hbox{\includegraphics[width=0.15\linewidth]{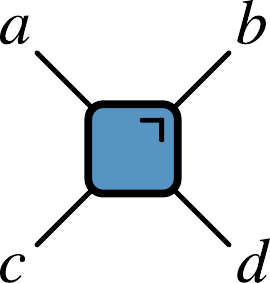}}}\,,\qquad U^{\dagger}_{ab,cd} =  \vcenter{\hbox{\includegraphics[width=0.15\linewidth]{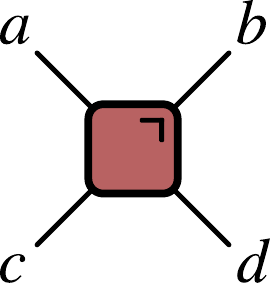}}}\,. 
\end{align}
In this notation each leg carries a local $q$-dimensional Hilbert space, not necessarily restricted to qubits with $q=2$, and the indices of legs connecting two operators are implicitly summed over (see e.g. Ref.~\cite{orus_practical_2014}). The unitarity of these gates can be graphically represented by
\begin{align}
\vcenter{\hbox{\includegraphics[width=0.65\linewidth]{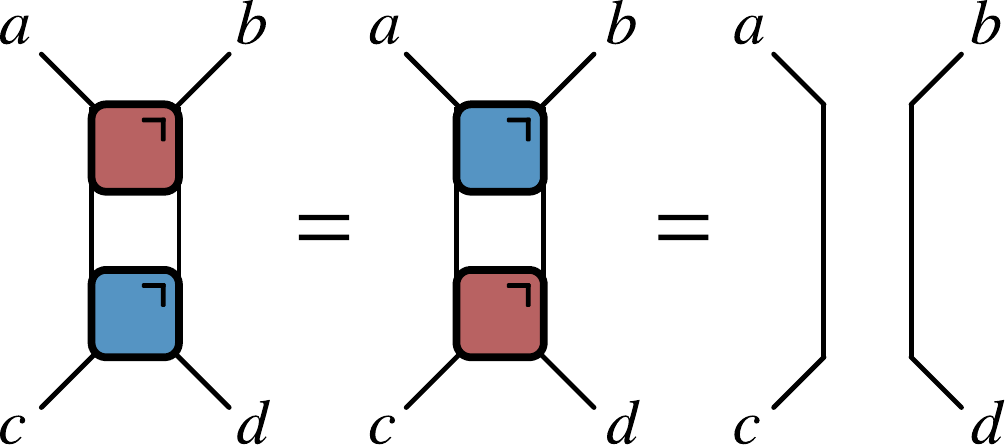}}}\,,
\end{align}
and labels for the outer legs are typically omitted unless this can generate confusion.

A unitary evolution operator acting on $N$ lattice sites can be constructed out of two-site unitary gates by considering a so-called `brick-wall' circuit, 
\begin{align}\label{eq:brickwall}
\mathcal{U}(t) =\, \vcenter{\hbox{\includegraphics[width=0.8\linewidth]{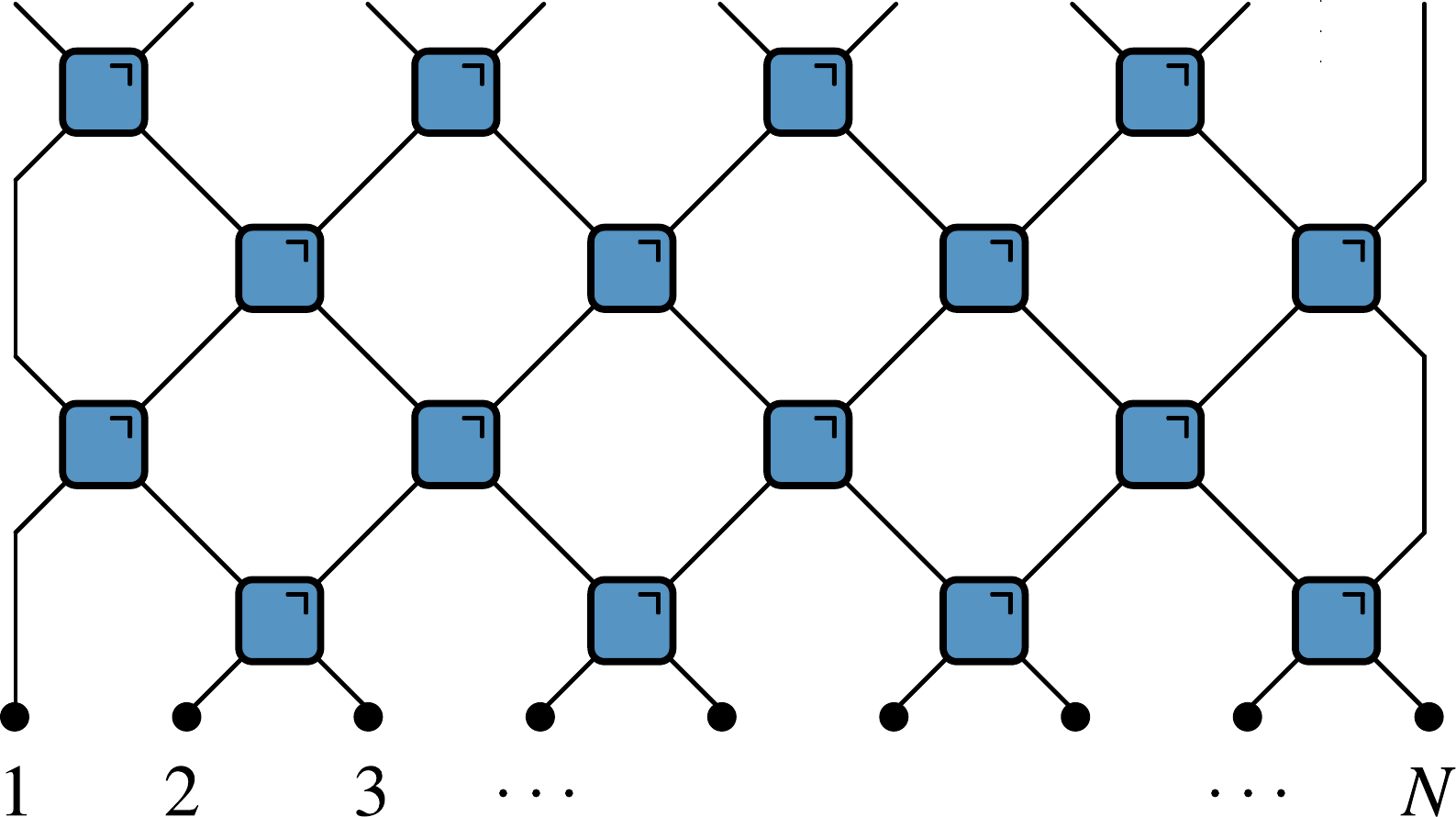}}}
\end{align}
The number of discrete time steps is fixed by the number of layers of unitary gates, where we choose a single time step to correspond to two layers, fixing the circuit to be periodic with period 1 such that the above graphic corresponds to $\mathcal{U}(t=2)$. Furthermore, we have closed boundary conditions and have chosen $N$ to be odd for concreteness. The former condition will turn out to be important, whereas all results can be directly extended to even $N$.

\section{State design in dual-unitary dynamics}
\label{sec:design_dualunitary}
Rather than directly calculating the $k$-th moments, we will introduce a replica-like trick and first calculate the related quantity
\begin{align}\label{eq:rho_nk}
\rho_{\mathcal{E}}^{(n,k)} = \sum_{z_B}p(z_B)^n \left(\ket{\tilde{\psi}(z_B)}\bra{\tilde{\psi}(z_B)}\right)^{\otimes k},
\end{align}
for $n \in \mathbbm{N}$ and where $\ket{\tilde{\psi}(z_B)} = \left(\mathbbm{1}_A \otimes \bra{z_B}\right)\ket{\psi}$ is the unnormalized wave function after projection. Unlike the case of the $k$-th moments, the averaging over measurement outcomes can be straightforwardly performed in this quantity, as will be shown below. The $k$-th moments of the projected ensemble are recovered for $n=1-k$. For all $n \in \mathbbm{N}$ we will show that $\rho_{\mathcal{E}}^{(n,k)} \propto\rho_{\textrm{Haar}}^{(k)}$, allowing us to directly extend the above result to negative $n=1-k$.

We partition the lattice in such a way that the leftmost $N_A$ sites correspond to the system sites and the remaining $N_B$ sites to bath sites. Graphically, $\ket{\tilde{\psi}(z_B)}\bra{\tilde{\psi}(z_B)}$ can be represented in the `folded' representation (see e.g. \cite{bertini_operator_2020}) as
\begin{align}\label{eq:operator_layer}
\vcenter{\hbox{\includegraphics[width=0.75\linewidth]{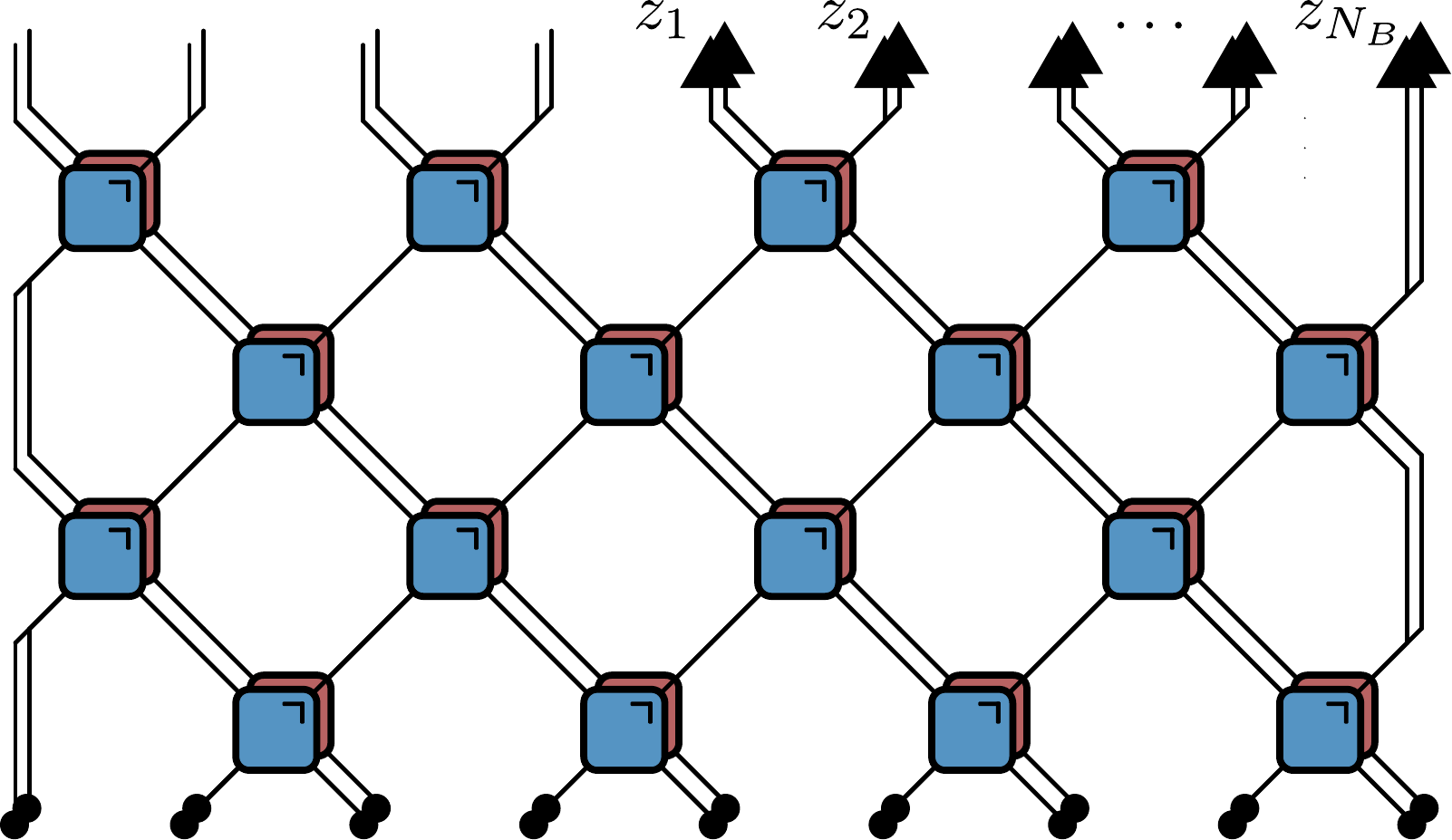}}}.
\end{align}
where we represent projectors $\ket{z}\bra{z}$ on the single-site measurement outcomes as triangles. For concreteness we also assume an initial product state, here indicated by black circles, representing e.g. $\ket{0}\bra{0}$. Tracing out the degrees of freedom on subsystem $A$, the corresponding probability can similarly be represented as
\begin{align}\label{eq:prob_layer}
\vcenter{\hbox{\includegraphics[width=0.75\linewidth]{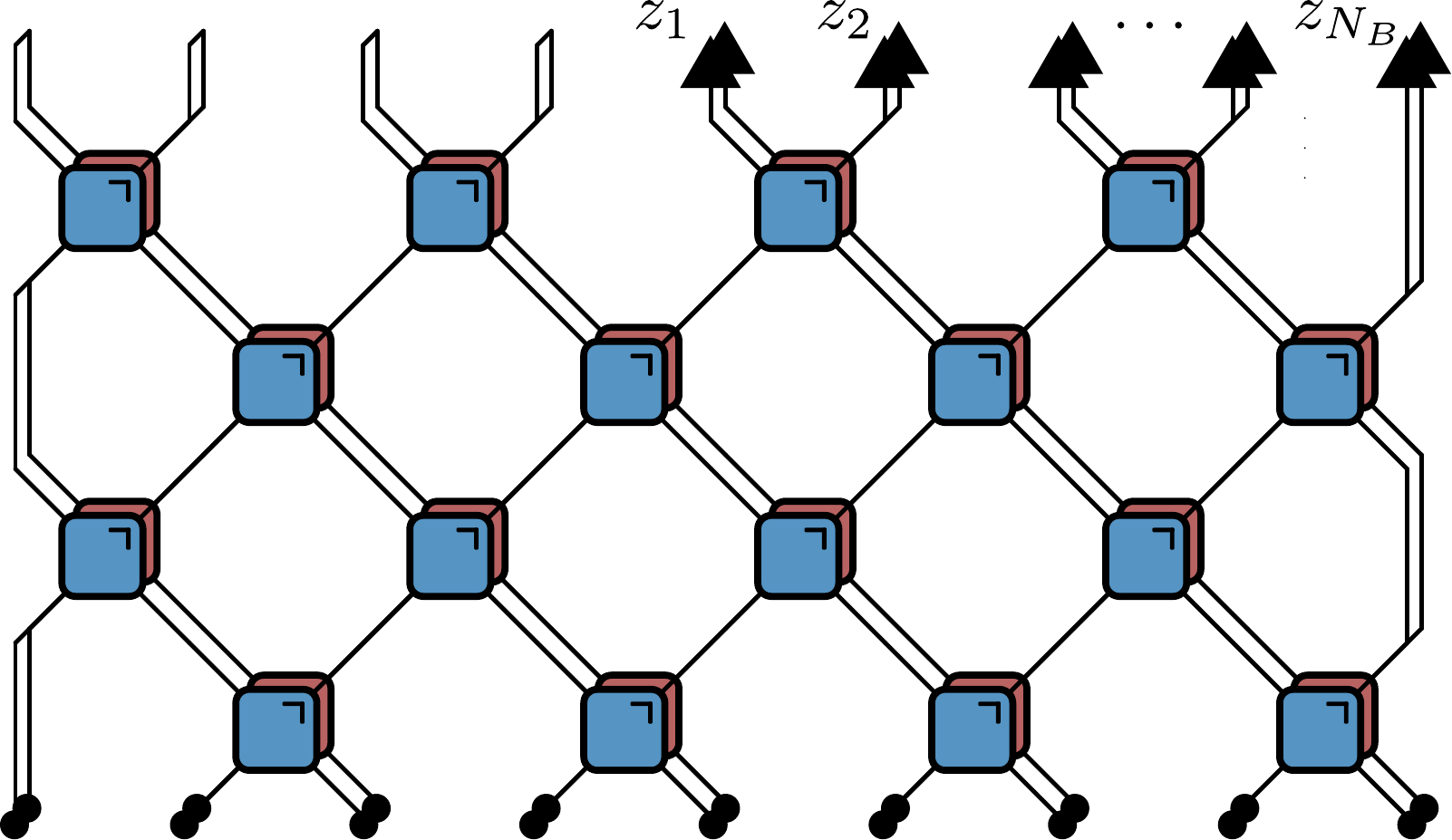}}}.
\end{align}
Eq.~\eqref{eq:rho_nk} then follows by stacking $k$ copies of Eq.~\eqref{eq:operator_layer} and $n$ copies of Eq.~\eqref{eq:prob_layer}. These different copies are related through the identical measurement outcomes in each layer, which are here grouped together. Averaging over the measurement outcomes, $\rho^{(n,k)}_{\mathcal{E}}$ is represented by
\begin{align}\label{eq:rho_nk_graph}
\vcenter{\hbox{\includegraphics[width=0.8\linewidth]{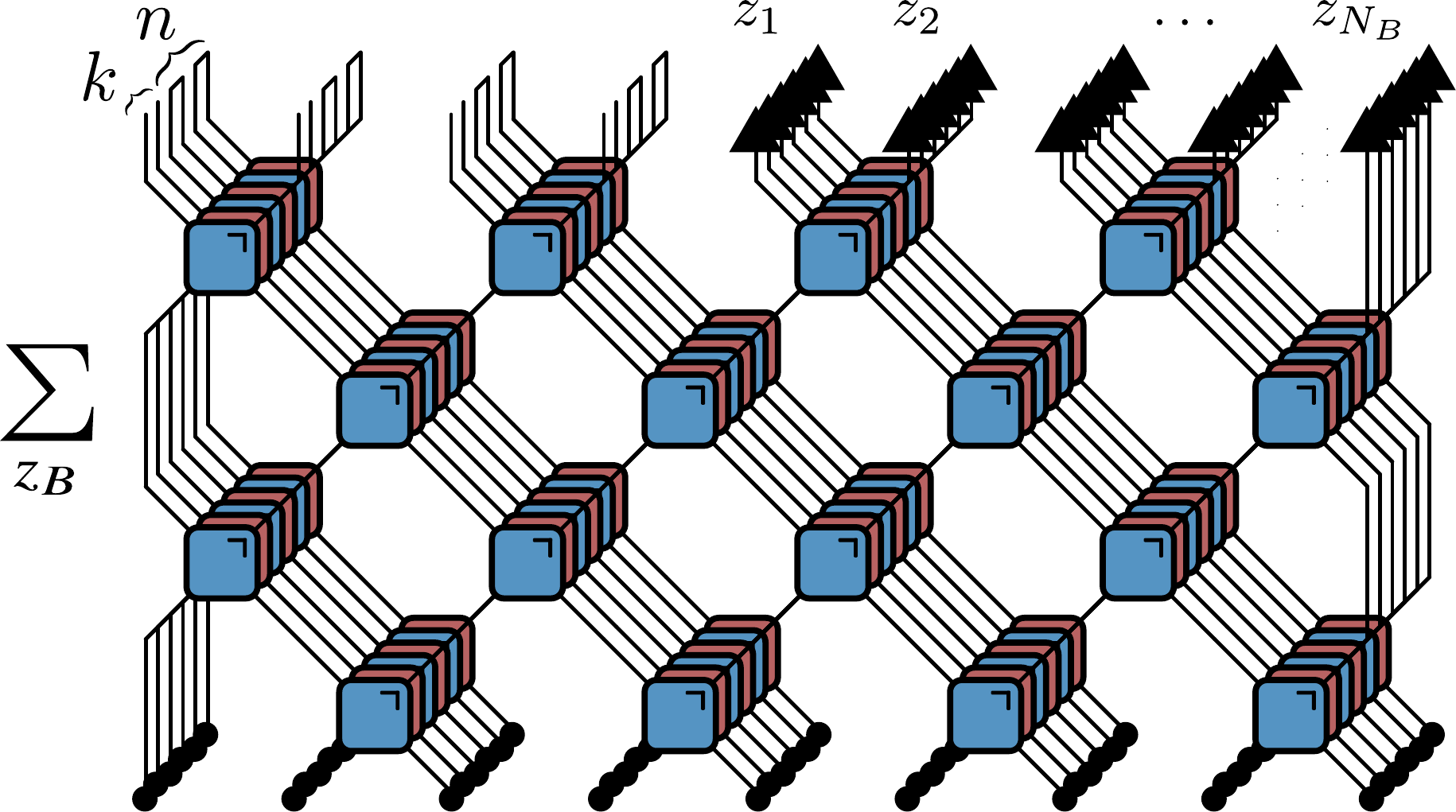}}}
\end{align}
To simplify notations we introduce a new notation for the folded gates consisting of $m=n+k$ copies of a pair of unitary and hermitian conjugate gates:
\begin{align}\label{eq:foldedgate}
\vcenter{\hbox{\includegraphics[width=0.7\linewidth]{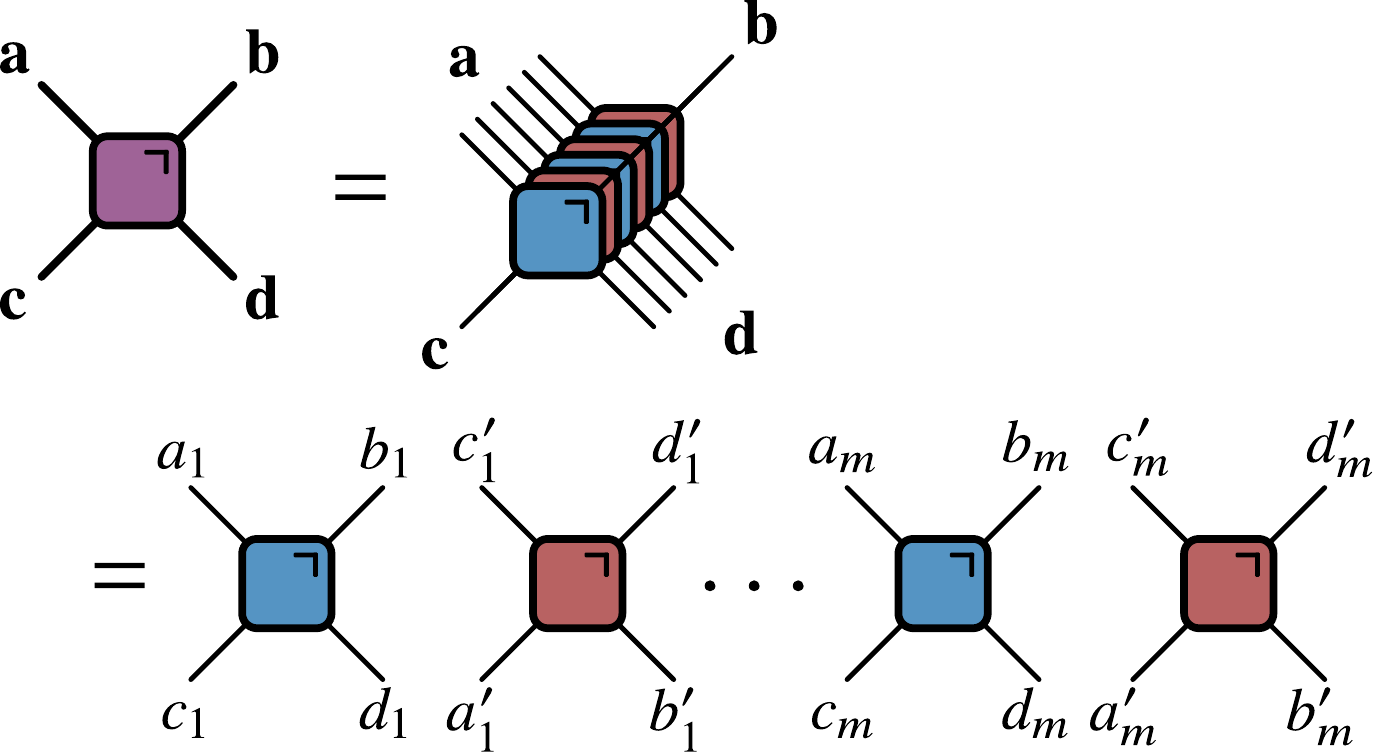}}}\,\,\,.
\end{align}
Each leg in this operator now acts on $2m$ copies of the local Hilbert space, which we index as $\textbf{a}=(a_1, a_1', a_2, a_2', \dots a_m, a_m')$ in order to distinguish the different hermitian conjugate layers.
The projectors on a measurement outcome $z_i$ can similarly be represented as
\begin{align}\label{eq:projectorz}
\vcenter{\hbox{\includegraphics[width=0.5\linewidth]{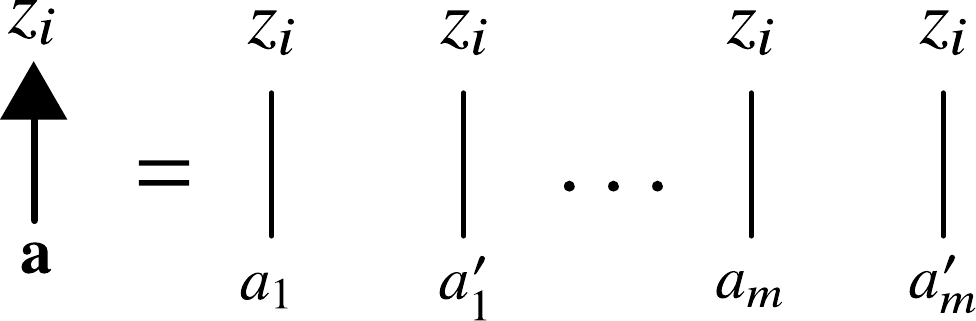}}}\,\,,
\end{align}
and the boundary condition tracing out the appropriate degrees of freedom on subsystem $A$ as
\begin{align}\label{eq:openbound}
\vcenter{\hbox{\includegraphics[width=0.9\linewidth]{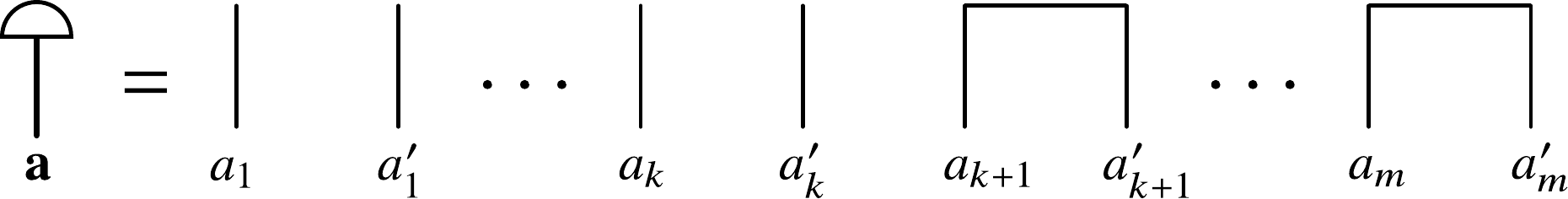}}}\,\,.
\end{align}
Taking this all together, the diagram from Eq.~\eqref{eq:rho_nk_graph} can be recast as
\begin{align}
\vcenter{\hbox{\includegraphics[width=0.85\linewidth]{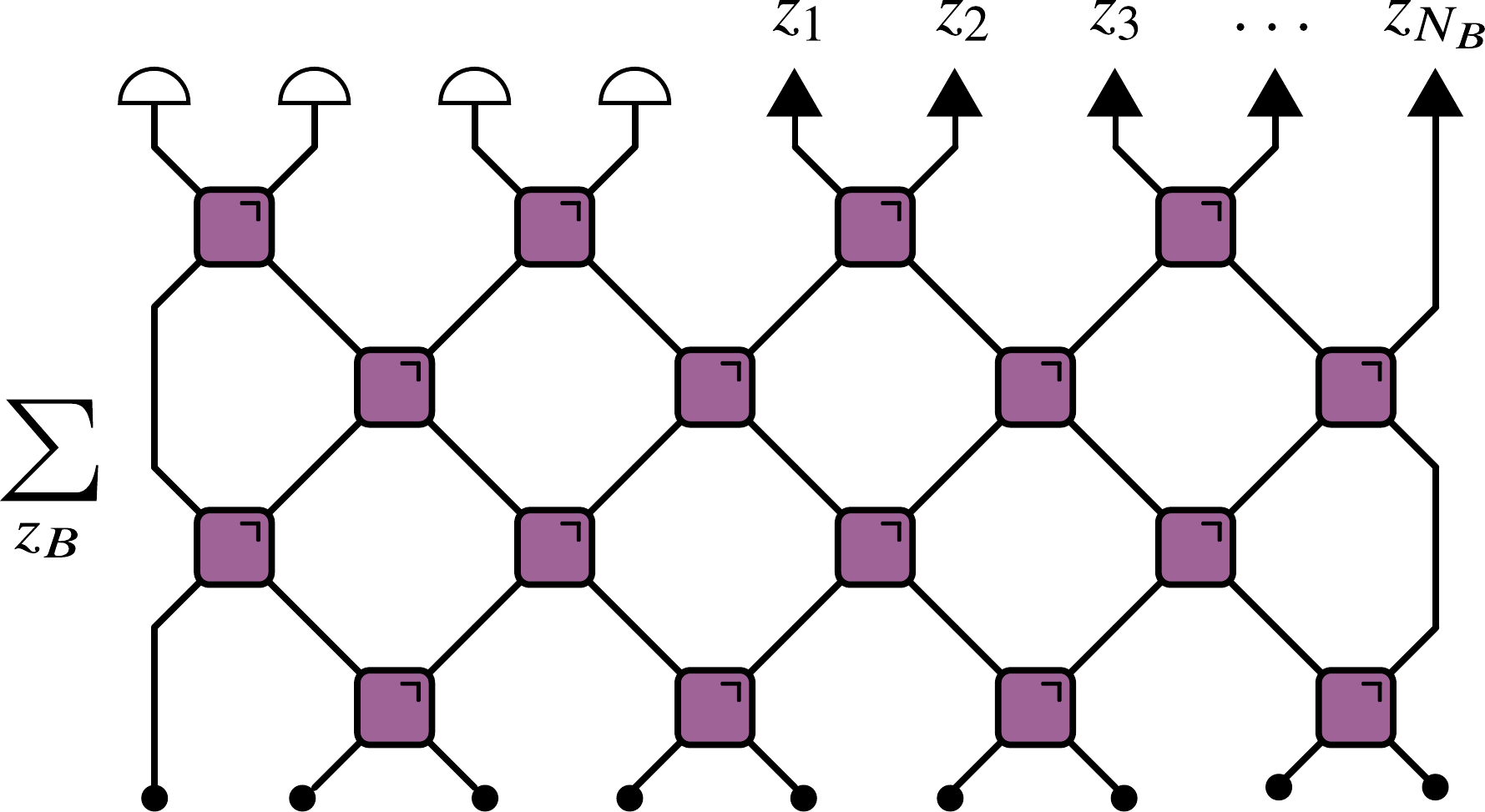}}},
\end{align}
This construction has the advantage that the averaging over measurement outcomes can be directly absorbed in the definition of a transfer matrix (similar to the construction from Ref.~\cite{ho_exact_2021}). This transfer matrix, here rotated by 90 degrees for convenience and illustrated for three time steps, is given by
\begin{align}\label{eq:transfermatrix}
\vcenter{\hbox{\includegraphics[width=0.75\linewidth]{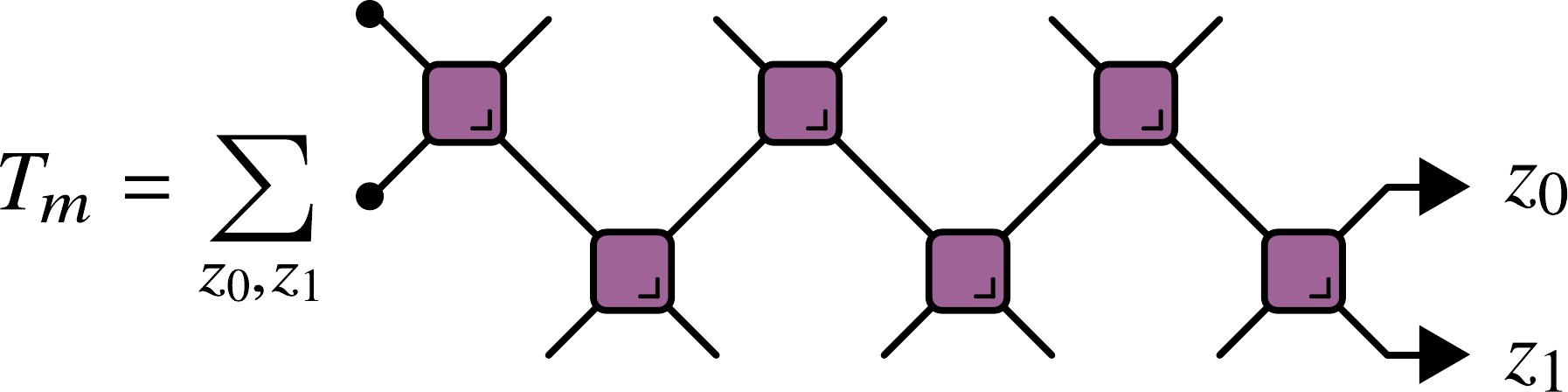}}}\,\,.
\end{align}
The effect of the bath is then to apply this transfer matrix $(N_B-1)/2$ times. In the thermodynamic limit of an infinite bath, this transfer matrix can be replaced by a projector on its leading eigenspace. So far, the construction applies to general unitary circuits. In the following subsections, we show how restricting the gates to be dual-unitary allows for the explicit construction of the leading eigenoperators of the transfer matrix and an explicit construction of the projector on the corresponding eigenspace for select initial states and measurement schemes.

\subsection{Dual-unitarity}
\label{subsec:dualunitarity}

We focus on the special case where $U$ is not just unitary but \emph{dual-unitary}. These circuits are defined as follows. For a unitary gate $U$ the so-called dual operator $\tilde{U}$ can be defined by reshuffling the indices as
\begin{align}
\tilde{U}_{ab,cd} = U_{db,ca}\,.
\end{align}
Dual-unitary circuits are circuits for which the underlying gates are chosen so that both $U$ and $\tilde{U}$ are unitary,
\begin{align}
U^{\dagger}U = U U^{\dagger}= \mathbbm{1},\quad \tilde{U}^{\dagger}\tilde{U} = \tilde{U}\tilde{U}^{\dagger} = \mathbbm{1}.
\end{align}
Graphically, this is represented by
\begin{align}
\vcenter{\hbox{\includegraphics[width=0.8\linewidth]{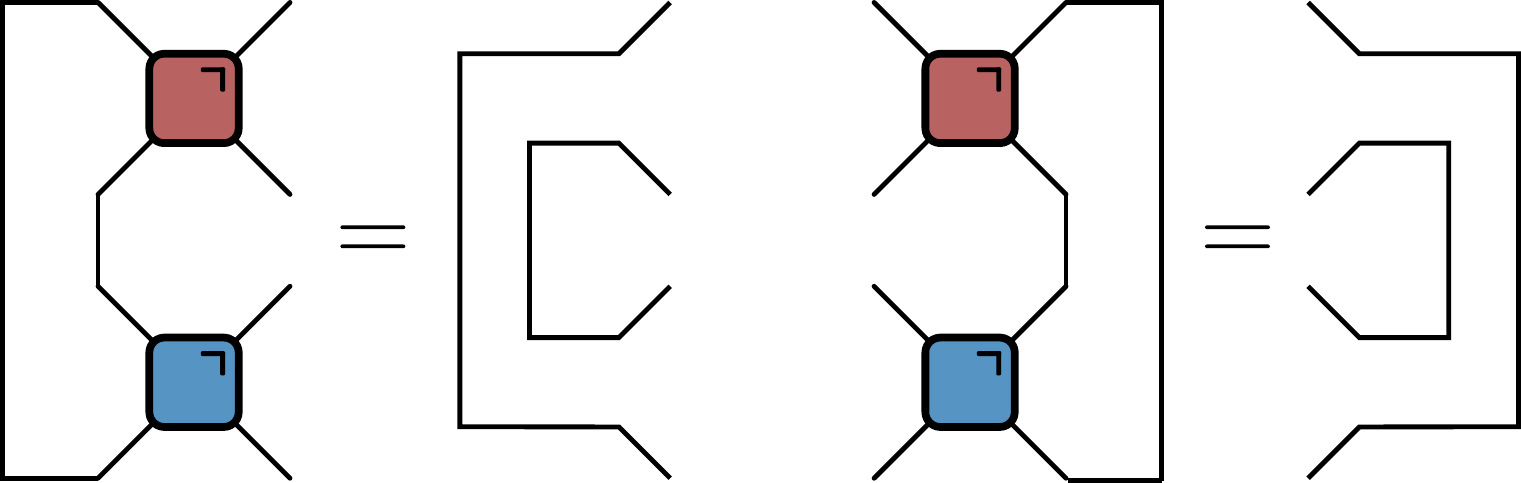}}}\,\,,
\end{align}

The construction of eigenstates of the transfer matrix \eqref{eq:transfermatrix} relies on the underlying dual-unitarity of these circuits. The transfer matrix $T_m$ contains $m$ layers of unitary gates and $m$ hermitian conjugate layers, such that there are $m!$ possible applications of dual-unitarity when choosing to contract a `unitary' layer with a hermitian conjugate one. This freedom can be illustrated by introducing operators
\begin{align}\label{eq:permutation_operators}
P(\pi_m)_{\textbf{a}}\, = \vcenter{\hbox{\includegraphics[width=0.7\linewidth]{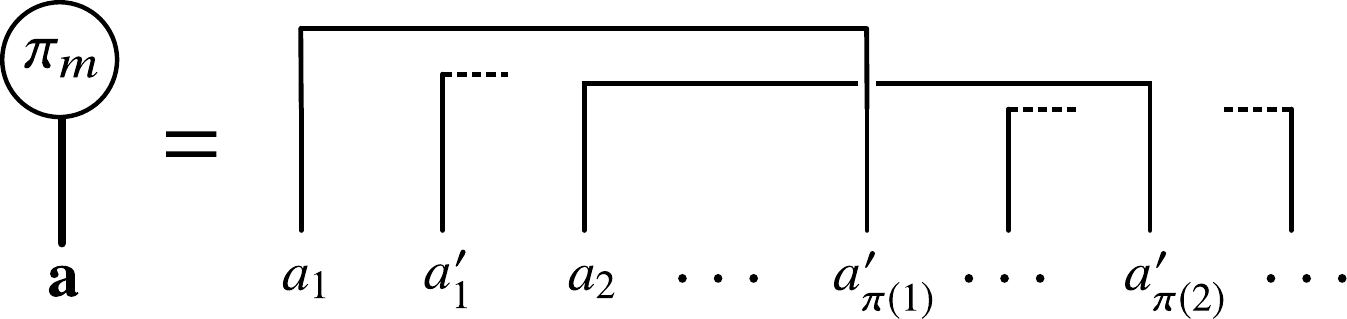}}},
\end{align}
labelled by the possible permutations $\pi \in S_m$ of $m$ elements (here effectively Hilbert spaces). Expressed in matrix elements, these are defined as
\begin{align}
P(\pi_m)_{\textbf{a}} = \delta_{a_1,a'_{\pi(1)}}\delta_{a_2,a'_{\pi(2)}}\dots\delta_{a_m,a'_{\pi(m)}}\,.
\end{align}
These operators directly correspond to the folded version of the permutation operators from Eq.~\eqref{eq:def_P_pi}, but acting on $m=n+k$ copies of the Hilbert spaces instead of $k$ copies. Such permutation operators can be used to express unitarity in the folded picture as
\begin{align}
\vcenter{\hbox{\includegraphics[width=0.85\linewidth]{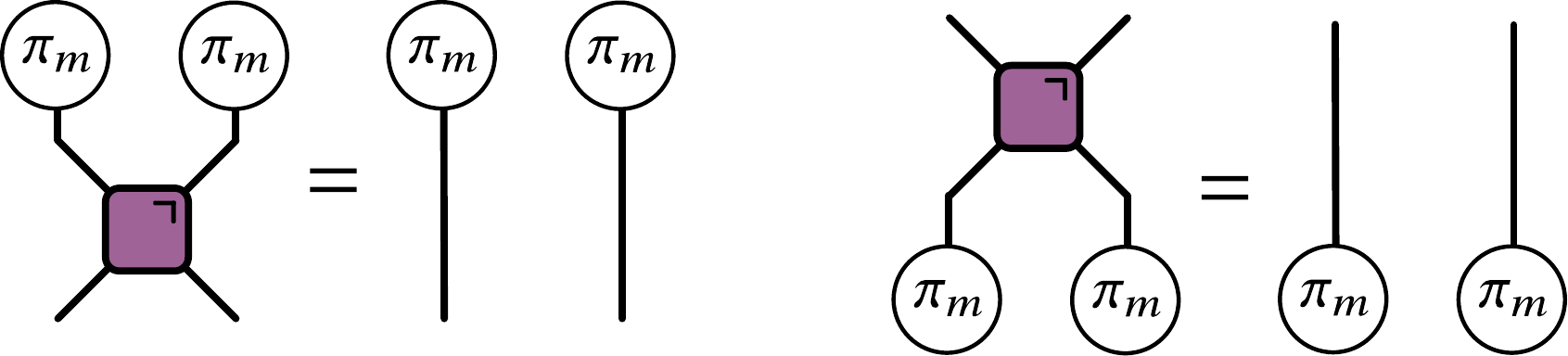}}}\,\,,
\end{align}
whereas dual-unitarity results in
\begin{align}
\vcenter{\hbox{\includegraphics[width=0.85\linewidth]{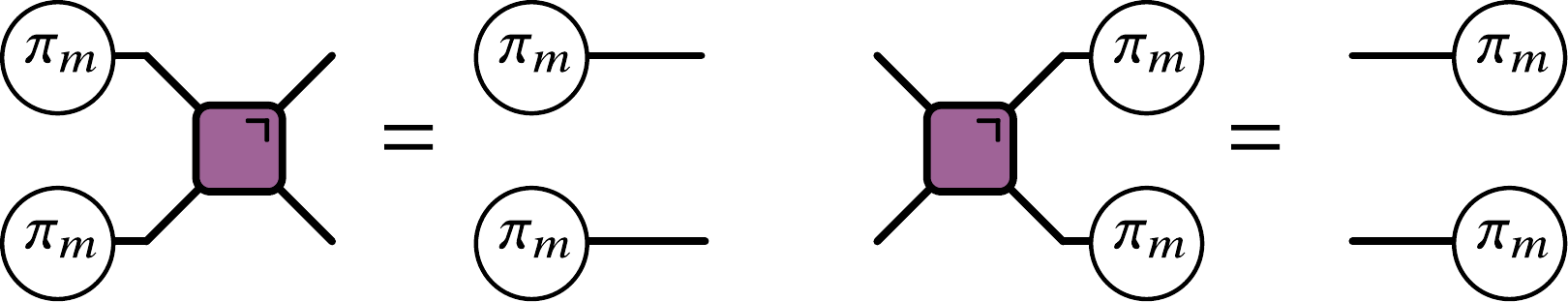}}}\,.
\end{align}
Acting with these operators on the transfer matrix \eqref{eq:transfermatrix}, we find that
\begin{align}\label{eq:partialeigen}
\vcenter{\hbox{\includegraphics[width=0.8\linewidth]{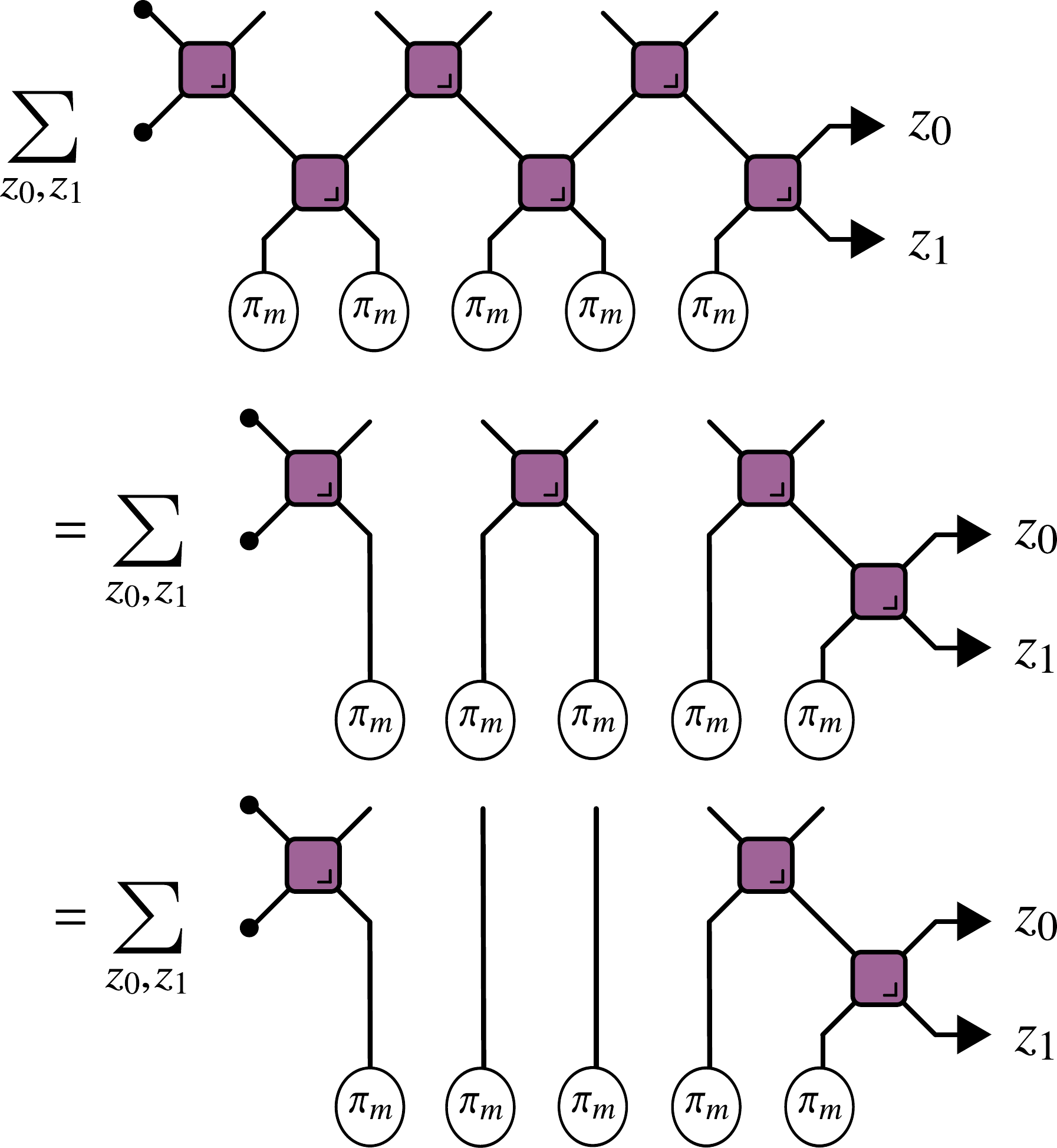}}}
\end{align}
Note again the rotation of the transfer matrix by 90 degrees, such that this equality depends on the dual-unitarity of the underlying gates rather than the unitarity. This expression suggests that eigenoperators can be constructed out of these permutation operators for dual-unitary circuits, provided additional constraints are satisfied at the boundaries -- here corresponding to the initial state and the projections on the measurement outcomes. In the following, we define a \emph{solvable measurement scheme} to be a measurement scheme for which the transfer matrix has a set of leading eigenoperators of the form
\begin{align}\label{eq:eigenstate_pi}
\vcenter{\hbox{\includegraphics[width=0.65\linewidth]{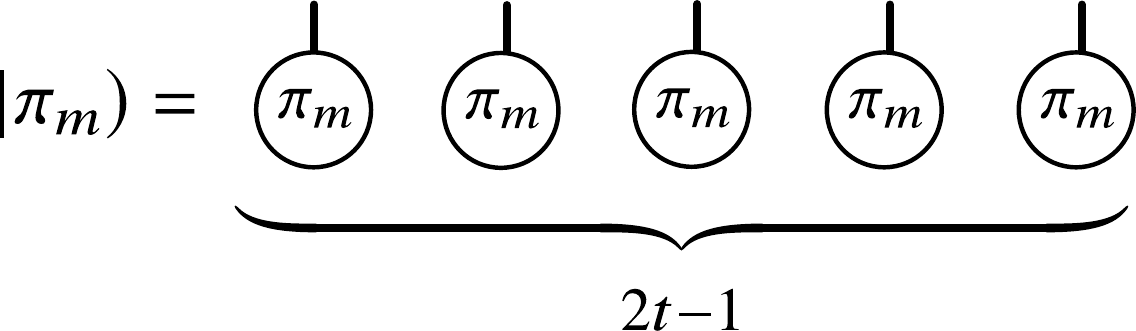}}}
\end{align}
Explicit examples of such solvable measurement schemes will be constructed in Sections \ref{subsec:Hadamard} and \ref{subsec:solvable_meas}.
In all presented results the transfer matrix for a fixed measurement outcome will be (proportional to) a unitary matrix. The averaging over different measurement outcomes is here equivalent to performing the Haar average over random unitary matrices, resulting in an emergent quantum state design. 

Let us show how a complete set of leading left and right eigenoperators of the form \eqref{eq:eigenstate_pi} suffices to lead to a quantum state design. These eigenoperators act as the permutation operators on $m$ copies of a $d_t=q^{2t-1}$ dimensional Hilbert space. While these operators do not form an orthonormal basis, results from random matrix theory can be applied to construct the projector on this eigenspace. Namely, if we take the Haar average of $m$ copies of a random unitary matrix and its Hermitian conjugate, each acting on a $d$-dimensional Hilbert space,
\begin{equation}
\includegraphics[width=0.7\linewidth]{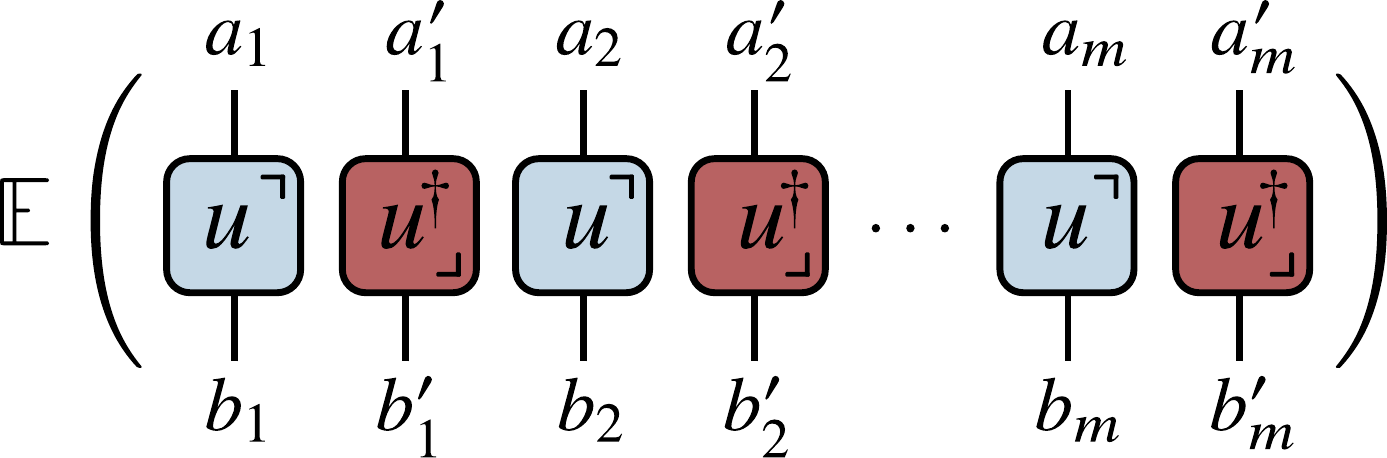}
\end{equation}
this average is given by
\begin{align}
\sum_{\sigma,\tau \in S_m} \delta_{a_1,a_{\sigma(1)}'}\dots \delta_{a_k,a_{\sigma(k)}'}&\delta_{b_1,b_{\tau(1)}'}\dots \delta_{b_m,b_{\tau(m)}'} \nonumber\\
&\times \textrm{Wg}(\sigma\tau^{-1},d)\,,
\end{align}
with Wg the Weingarten functions \cite{weingarten_asymptotic_1978,collins_moments_2003,collins_integration_2006}. As follows from its definition as a Haar average, this operator is a projector and its eigenoperators are exactly the permutation operators $P(\pi_m)$ acting on a $d$-dimensional Hilbert space. For the eigenoperators of Eq.~\eqref{eq:eigenstate_pi}, this implies that we can write the projector on the corresponding eigenspace as
\begin{align}
\sum_{\sigma_m,\tau_m \in S_m} \textrm{Wg}(\sigma_m \tau_m^{-1},d=q^{2t-1}) \, |\tau_m)(\sigma_m|\,,
\end{align}
or, graphically,
\begin{align}
\includegraphics[width=0.9\linewidth]{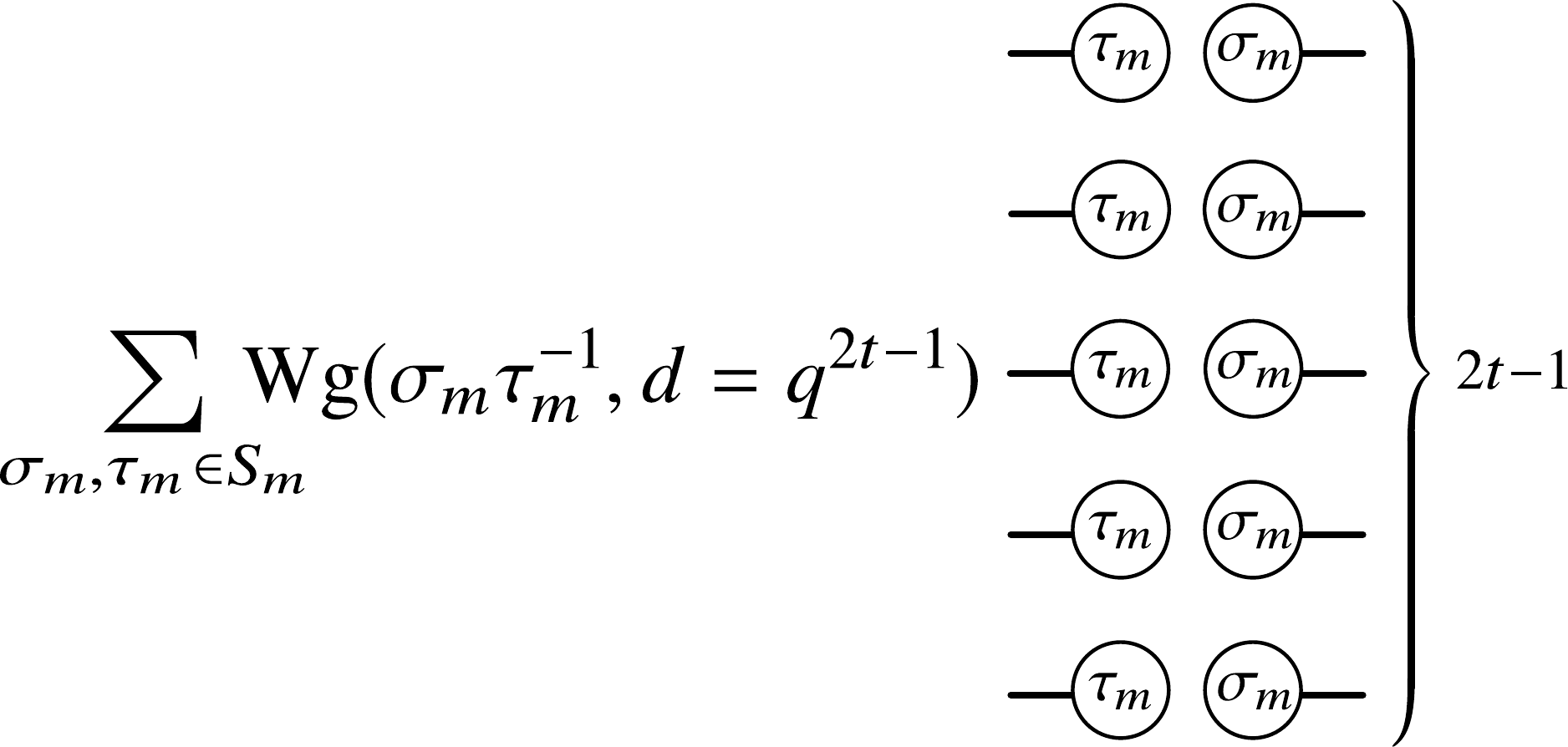}
\end{align}
In other words, the averaging over different measurement outcomes is equivalent to sampling from Haar-random unitaries acting along the space direction.
Inserting this projector in the expression for $\rho^{(n,k)}_{\mathcal{E}}$, the action on both boundaries can be evaluated in a relatively straightforward way (see Appendix~\ref{app:boundaries}). The contraction with the right boundary returns a constant, whereas the left boundary maps the contractions along the space direction to contractions in the physical Hilbert space while preserving the random matrix structure (provided $t>N_A/2$). Combining these results, we find that
\begin{align}
&\lim_{N_B \to \infty} q^{2(m-1)N_B} \sum_{z_B} p(z_B)^n  \left(\ket{\tilde{\psi}(z_B)}\bra{\tilde{\psi}(z_B)}\right)^{\otimes k} \nonumber\\
&\quad\propto \sum_{\pi_k \in S_k}P(\pi_k)\,,
\end{align}
where the prefactor is not complicated but rather unwieldy and is given in Appendix \ref{app:boundaries}. The factor $q^{2(m-1)N_B}$ is required to obtain a finite and nonzero result in the limit $N_B \to \infty$, but vanishes for $n=1-k$ and hence $m=1$. This result can be extended to $n=1-k$ by analytic continuation since the operator part $\propto \rho_{\textrm{Haar}}^{(k)}$ does not depend on $n$ and the prefactor can be directly evaluated for $m=1$ to return
\begin{align}
\rho^{(k)}_{\mathcal{E}} = \rho^{(k)}_{\textrm{Haar}}\,,\qquad \forall k. 
\end{align}
For $t>N_A/2$ the projected ensemble returns an exact quantum state design in the limit of an infinite bath. Having established that our definition of solvable measurement schemes results in an exact emergent quantum state design, we will now consider two families of dual-unitary dynamics where this criterion can be satisfied.

\subsection{Complex Hadamard gates}
\label{subsec:Hadamard}

We first consider single-site measurements in the computational basis, as a direct extension of the results of Ref.~\cite{ho_exact_2021} for the kicked Ising model (KIM).
The KIM takes a special place within the class of dual-unitary gates.
In the KIM unitary evolution $e^{-iH_{I}}$ with a classical Hamiltonian $H_I = J\sum_{j}Z_{j}Z_{j+1}+\sum_{j}h_j Z_j$ is alternated with a kick operator $K = e^{-iH_K}$, $H_K = b\sum_j X_j$, where $X_j$ and $Z_j$ are Pauli matrices acting on site $j$. This Hamiltonian dynamics can be mapped to a unitary circuit of the form \eqref{eq:brickwall} by writing
\begin{align}\label{eq:DFTgate}
U = \mathcal{I} (\mathcal{K} \otimes \mathcal{K}) \mathcal{I}  = \vcenter{\hbox{\includegraphics[width=0.16\linewidth]{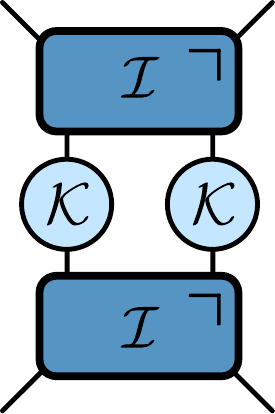}}}\, .
\end{align}
Here $\mathcal{I}_{12} = \exp\left[iJ\left(Z_{1}Z_{2}\right)+i(h_1 Z_1+h_2 Z_2)/2\right] $ and $\mathcal{K}_{1,2}=\exp[ibX_{1,2}]$, with $X$ and $Z$ Pauli matrices. For $|J|=|b| = \pi/4$ these gates are dual-unitary. 

All dual-unitary gates for a local two-dimensional Hilbert space were fully classified in Ref.~\cite{bertini_exact_2019}, and various extensions of the KIM to higher-dimensional Hilbert spaces have already been proposed (see Refs.~\cite{gutkin_exact_2020,aravinda_dual-unitary_2021}).
Here we introduce a class of dual-unitary gates for arbitrary local Hilbert space dimension $q$ built out of four (possibly different) complex Hadamard matrices, encompassing all such previous constructions and highlighting their shared properties. This construction depends on the biunitarity of the Hadamard matrices, as detailed in Appendix~\ref{app:biunitary}. A complex Hadamard matrix $H \in \mathbbm{C}^{q\times q}$ satisfies
\begin{align}
&H^{\dagger}H = H H^{\dagger} = q \mathbbm{1}, \nonumber\\
& |H_{ab}|=1,\quad \forall a,b=1\dots q.
\end{align}
Complex Hadamard matrices are proportional to unitary matrices, and all matrix elements have modulus one. This additional property on top of the unitarity underlies their \emph{biunitarity} \cite{reutter_biunitary_2019}. We can construct a two-site dual-unitary gate out of four complex Hadamard matrices $E,F,G,H \in \mathbbm{C}^{q\times q}$ as
\begin{align}\label{eq:Hadamardgate}
U_{ab,cd} = E_{ab}F_{bd}G_{dc}H_{ca}/q\,.
\end{align}
Note that the dual of $U$ can be written in the exact same way, i.e.
\begin{align}
\tilde{U}_{ab,cd}=U_{db,ca} = F^T_{ab}E^T_{bd}H^T_{dc}G^T_{ca}/q\,,
\end{align}
since the transpose of a complex Hadamard matrix is again a complex Hadamard matrix. Due to this symmetry unitarity here immediately implies dual-unitarity. Both unitarity and dual-unitarity can be proven by direct calculation, and this parametrization encompasses previous constructions incorporating Hadamard matrices \cite{gutkin_exact_2020} and `cat maps' \cite{aravinda_dual-unitary_2021}. Note that previous generalizations of dual-unitary gates based on Fourier matrices, a specific class of complex Hadamard matrices, were also introduced in Ref.~\cite{borsi_remarks_2022}.

Next to their dual-unitary, these models share the special property of the KIM that product states in the computational basis effectively behave as `solvable states'. Namely, for these unitary gates
\begin{align}\label{eq:KIM_identity}
\vcenter{\hbox{\includegraphics[width=0.6\linewidth]{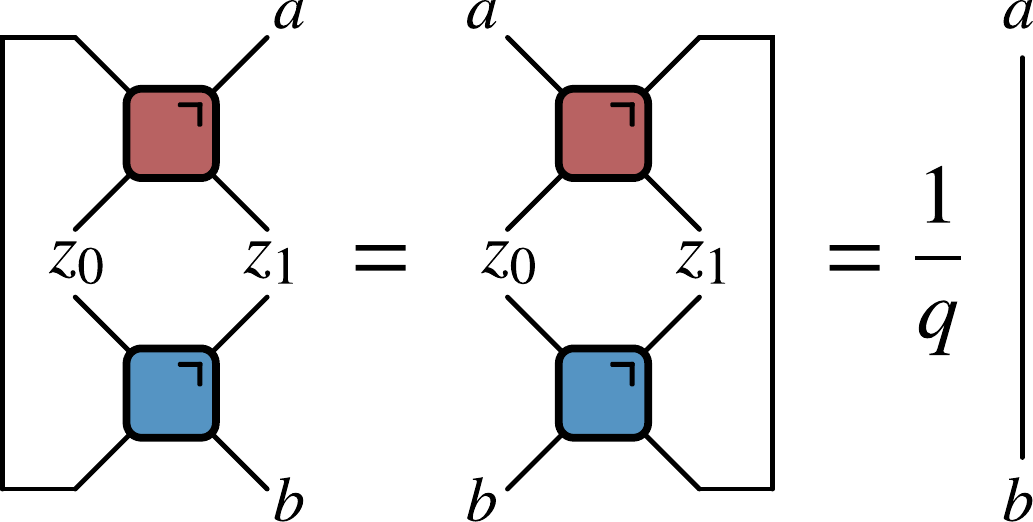}}}\,\,\,,
\end{align}
for any fixed $z_0,z_1$. This equality can be directly verified by plugging in the parametrization \eqref{eq:Hadamardgate}. Expressed in matrix elements, this identity reads
\begin{align}
\sum_{c} U_{z_0z_1,cb} U^{\dagger}_{ca,z_0z_1} =\sum_{c} U_{z_0z_1,bc} U^{\dagger}_{ac,z_0z_1}= \frac{1}{q} \delta_{ab}\,.
\end{align}
This property was first observed for the KIM in Ref.~\cite{gopalakrishnan_unitary_2019}, where it was used to construct the reduced density matrix for a system initially prepared in a product state and evolved using the KIM. As far as we are aware, its extension to complex Hadamard gates and cat maps has not been previously observed, but futher strengthens the interpretation of these circuits as higher-dimensional KIMs. This equality has the direct interpretation that these gates map two-site product states in the computational basis to maximally entangled states, since the reduced density matrix on a single site is proportional to the identity. Maximally entangled states play a special role in the theory of dual-unitarity, as will be discussed in Sec.~\ref{subsec:solvable_meas}.

Both the dual-unitarity and the KIM property \eqref{eq:KIM_identity} are preserved under the addition of one-site unitary gates that are diagonal in the computational basis, i.e. we can consider dual-unitary gates of the form
\begin{align}
U_{ab,cd} &= E_{ab}F_{bd}G_{dc}H_{ca}/q \nonumber\\
&\times\exp\left[\frac{i}{2}(h_1[a]+h_2[b]+h_3[c]+h_4[d])\right]\,,
\end{align}
for arbitrary real functions $h_{i}[a],i=1\dots 4$. For the circuit dynamics only $h_1+h_4$ and $h_2+h_3$ are relevant, such that we can take $h_3=h_4=0$ without loss of generality. For $q=2$ the KIM can be written in this form and for general $q$ with $E=G$ and $F=H$ these circuits again have an interpretion as a periodically kicked chain (see Ref.~\cite{gutkin_exact_2020} and Appendix \ref{app:biunitary}). 

Taking $m$ copies of the identity \eqref{eq:KIM_identity}, we find for the folded gates of Eq.~\eqref{eq:foldedgate} and the permutation operators of Eq.~\eqref{eq:permutation_operators} the graphical identity
\begin{align}\label{eq:KIM_identity_layered}
\vcenter{\hbox{\includegraphics[width=0.65\linewidth]{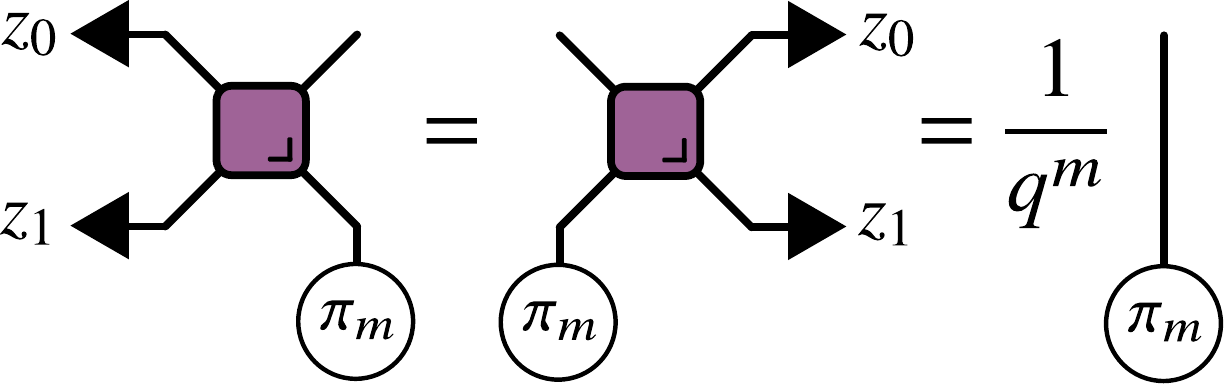}}}\,.
\end{align}
This identity combined with the dual-unitarity suffices to find a set of eigenoperators for the transfer matrix $T_m$ of the form Eq.~\eqref{eq:eigenstate_pi}, i.e.
\begin{align}\label{eq:eigenstate_pi_2}
\vcenter{\hbox{\includegraphics[width=0.6\linewidth]{fig_eigenstate_pi}}}\,\,,
\end{align}
with corresponding eigenvalue $q^{2(1-m)}$. These necessarily correspond to the leading eigenvalues of $T_m$. Each of the individual transfer matrices for fixed $z_0,z_1$ equals a unitary matrix times $q^{-2m}$, as can be shown by again combining dual-unitarity with Eq.~\eqref{eq:KIM_identity}. In other words, each of the eigenvalues of a transfer matrix with fixed $z_0,z_1$ has modulus $q^{-2m}$. The eigenoperators constructed out of permutation operators are common eigenoperators of all transfer matrices with fixed $z_0,z_1$ and hence also of the total transfer matrix $T_m$, with eigenvalue $q^2 \times q^{-2m}$, which is an upper bound to the eigenvalues of this transfer matrix. This does not guarantee that these eigenoperators exhaust the leading eigenoperators (which indeed isn't the case at the integrable point), but numerical checks for small $m$ and $t$ suggest that these are in fact exhaustive away from the integrable point. Following the arguments from Sec.~\ref{subsec:dualunitarity}, we conclude that the combination of complex Hadamard gates with an initial state and measurements in the computational basis leads to an exact quantum state design for $t>N_A/2$.

Before moving on to more general measurement schemes, we note that this construction fails at $h_1=h_2=0$. At this point the KIM is integrable and the dynamics are purely solitonic \cite{bertini_operator_ii_2020}. In our scheme, the integrability is reflected in the appearance of additional leading eigenoperators. Alternatively, this lack of emergent randomness can be seen to follow directly from the gates: at the integrable point the KIM gate is a Clifford gate, such that the unitary dynamics is given by a Clifford circuit. For all presented setups the Clifford dynamics can be efficiently simulated on a classical computer, similar to the typical solvability of quantum integrable models. Crucially, Clifford circuits do not form a universal gate set and form a unitary $3$-design \cite{webb_clifford_2016}, such that no emergent state design can be expected. Similar behaviour is observed for general complex Hadamard gates, although a full analysis falls outside the scope of this paper. Note that the reduced density matrix here always thermalizes to infinite temperature, even at the integrable point (see e.g. \cite{gopalakrishnan_unitary_2019}), but this does not imply an emergent state design -- in the language of Ref.~\cite{wilming_high-temperature_2022}, at the integrable point the computational basis remains a highly atypical measurement basis at all times.

\subsection{Solvable MPS measurement schemes}
\label{subsec:solvable_meas}

So far, only single-site measurements in the computational basis were considered, where the construction from Sec.~\ref{subsec:dualunitarity} imposes additional constraints on the dual-unitary gates in order to lead to an emergent quantum state design. An alternative solvable measurement scheme can be introduced based on solvable initial matrix product states (MPSs), avoiding the explicit dependence on the unitary gates through identities of the form \eqref{eq:KIM_identity}.
In Ref.~\cite{piroli_exact_2020} Piroli and Bertini introduced a family of MPSs for which generic dual-unitary circuit dynamics remains solvable. These states are written in 2-site MPS form as
\begin{align}\label{eq:def_MPS}
\ket{\psi_0} = \sum_{i_1,i_2,\dots,i_N}\tr\left[\mathcal{N}^{(i_1,i_2)}\mathcal{N}^{(i_3,i_4)}\dots\right]\ket{i_1 i_2 i_3\dots }\,,
\end{align}
where $\{\mathcal{N}^{(i,j)}\}_{i,j=1}^{q}$ is a set of $\chi$-dimensional matrices. Here $\chi$ is the bond dimension of the MPS, and these matrices can again be graphically represented as
\begin{align}\label{eq:MPS_N}
\includegraphics[width=0.5\linewidth]{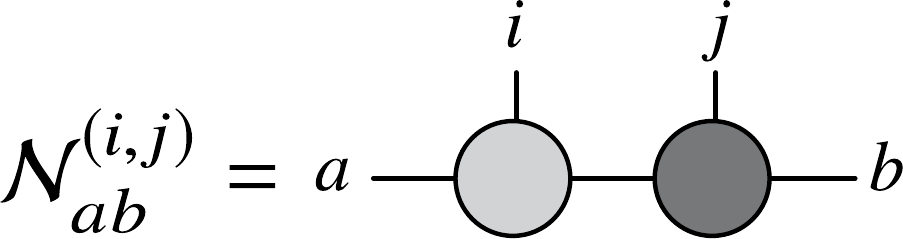}
\end{align}
The underlying solvability is highly similar to the definition of dual-unitarity. First define a matrix $\mathcal{W}(\mathcal{N})$ acting on the tensor product of the physical space and the auxiliary (bond dimension) space as
\begin{align}
\bra{i}\otimes\bra{a} \mathcal{W}(\mathcal{N}) \ket{j} \otimes \ket{b} = \sqrt{q}\, \mathcal{N}^{(i,j)}_{ab} \,.
\end{align}
This matrix can be interpreted as evolving the state from the right to the left [c.f. Eq.~\eqref{eq:MPS_N}]. The state is solvable if $\mathcal{W}(\mathcal{N})$ is unitary, resulting in the graphical identities
\begin{align}\label{eq:solvabilityMPS}
\includegraphics[width=0.45\linewidth]{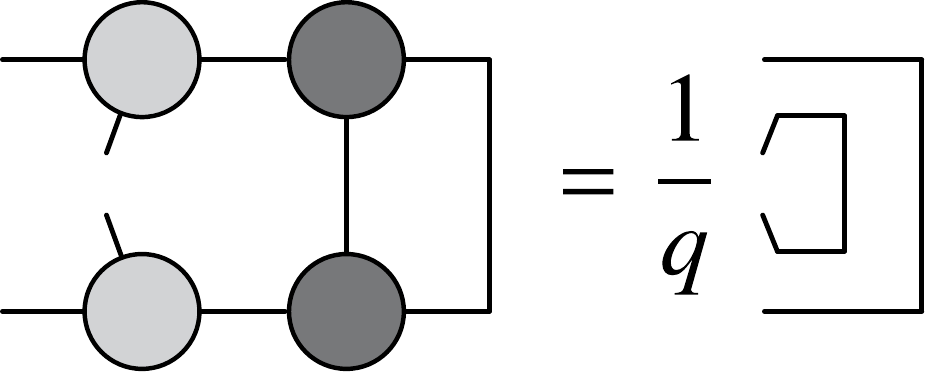} \\
\includegraphics[width=0.45\linewidth]{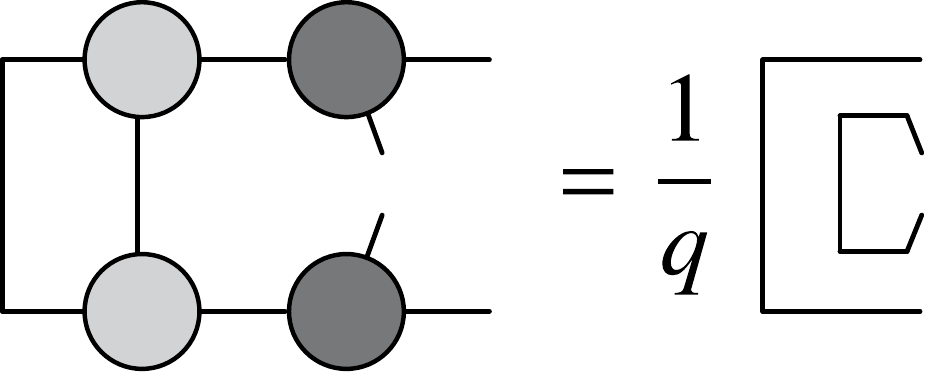}
\end{align}
Note the similarity with Eq.~\eqref{eq:KIM_identity}. This property was taken as the defining property of solvability for an MPS, and different classes of solvable initial states were explicitly constructed \cite{piroli_exact_2020}. Taking the initial state of the dynamics to be a solvable MPS can be used to contract the initial state contribution to Eq.~\eqref{eq:partialeigen}.

However, the transfer matrix construction for the projected ensemble depends on both the initial state and the measurement scheme. If the projection operators on the measurement outcomes can be expressed as a solvable MPS, the construction can be directly extended, but this is far from guaranteed. Such projectors need to form a complete basis, and we furthermore wish to have purely local measurements -- similar in spirit to the single-site measurements of the previous section.
Here, we present a measurement scheme for which the measurements give rise to an exact quantum state design for any initial solvable state and ergodic dual-unitary circuit dynamics.  We will first illustrate this construction for a local two-dimensional Hilbert space $q=2$ (qubits), before presenting extensions to higher-dimensional Hilbert spaces.

Consider measurements on two neighbouring sites in the Bell pair basis, i.e. 
\begin{align}
\ket{\phi^{\pm}} &= \frac{1}{\sqrt{2}}\left(\ket{00}\pm\ket{11}\right),\nonumber\\
\ket{\psi^{\pm}} &= \frac{1}{\sqrt{2}}\left(\ket{01}\pm\ket{10}\right)\,.
\end{align}
These form a complete basis, and the corresponding states can be represented as
\begin{align}
\includegraphics[width=0.26\linewidth]{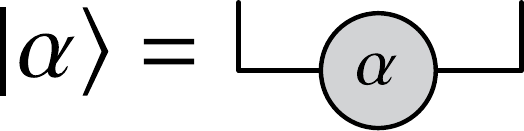}
\end{align}
with $\alpha \in \{\phi^{\pm},\psi^{\pm}\}$ and the $q\times q$ matrix labelled by $\alpha$ defined through its matrix elements as $\left(\alpha\right)_{ij}= \braket{ij|\alpha}$. A product state consisting of Bell pair states can be written as an MPS with
\begin{align}
\mathcal{N}^{(i,j)} = \alpha_{ij}=\braket{ij|\alpha}\,.
\end{align}
This expression returns the simplest example of a solvable MPS, with bond dimension $1$. The solvability of the MPS is underlied by the unitarity of each $\alpha$, since $\alpha^{\dagger}\alpha = \alpha \alpha^{\dagger} = \mathbbm{1}/q$ and hence
\begin{align}\label{eq:solvablepairstate}
\includegraphics[width=0.65\linewidth]{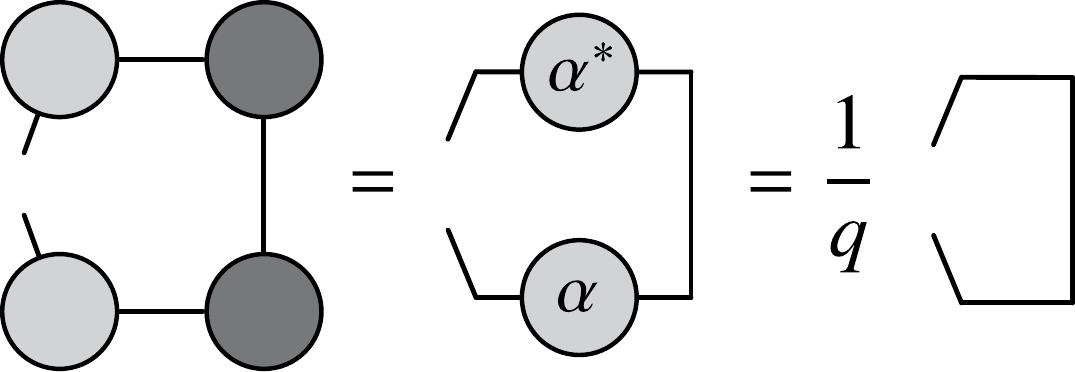}
\end{align}
When comparing this expression with Eq.~\eqref{eq:solvabilityMPS}, it is useful to note that the lines corresponding to a 1-dimensional Hilbert space are not drawn.
While the above discussion focused on measurements in a Bell pair basis and hence for qubits, a similar measurement basis can be found for arbitrary dimensions of the Hilbert space. Measuring pairs of neighbouring sites, the above construction required (i) that the set of states $\ket{\alpha}$ forms a complete basis (since these follow from a measurement scheme) and (ii) the corresponding $q\times q$ matrices $\alpha$ are unitary and hence give rise to a solvable MPS. For arbitrary dimensions of the local Hilbert space these two requirements are satisfied if the matrices $\alpha$ form a \emph{unitary error basis} \cite{knill_non-binary_1996}. A unitary error basis is a collection of unitary matrices $\{\alpha_n,n=1\dots q^2\}$ satisfying the orthogonality property
\begin{align}
\tr\left(\alpha_n^{\dagger}\alpha_m\right) = q\delta_{m,n}\,.
\end{align}
Because of this orthonormality the corresponding states form a complete basis for the $q^2$-dimensional two-site Hilbert space, and the unitarity leads to a solvable MPS for any measurement outcome/value of $\alpha_n$. We say that unitary error bases lead to a \emph{solvable measurement scheme}. Unitary error bases are ubiquitous in quantum information since they are crucial for quantum teleportation, dense coding and error correction procedures \cite{knill_non-binary_1996,shor_fault-tolerant_1996,werner_all_2001}, and various schemes have been proposed to systematically realize unitary error bases \cite{reutter_biunitary_2019}. Note that a unitary error basis is a biunitary construction, same as complex Hadamard gates and dual-unitary circuits, further strengthening the interplay between these different constructions.

We can now consider a setup where the system is initally prepared in a solvable MPS of the form \eqref{eq:def_MPS}, evolved using dual-unitary circuit dynamics, and the $N_B$ bath states are measured in a solvable basis. The resulting circuit for $\rho_{\mathcal{E}}^{(n,k)}$ and hence the emergent quantum state design follows as
\begin{align}
\includegraphics[width=0.85\linewidth]{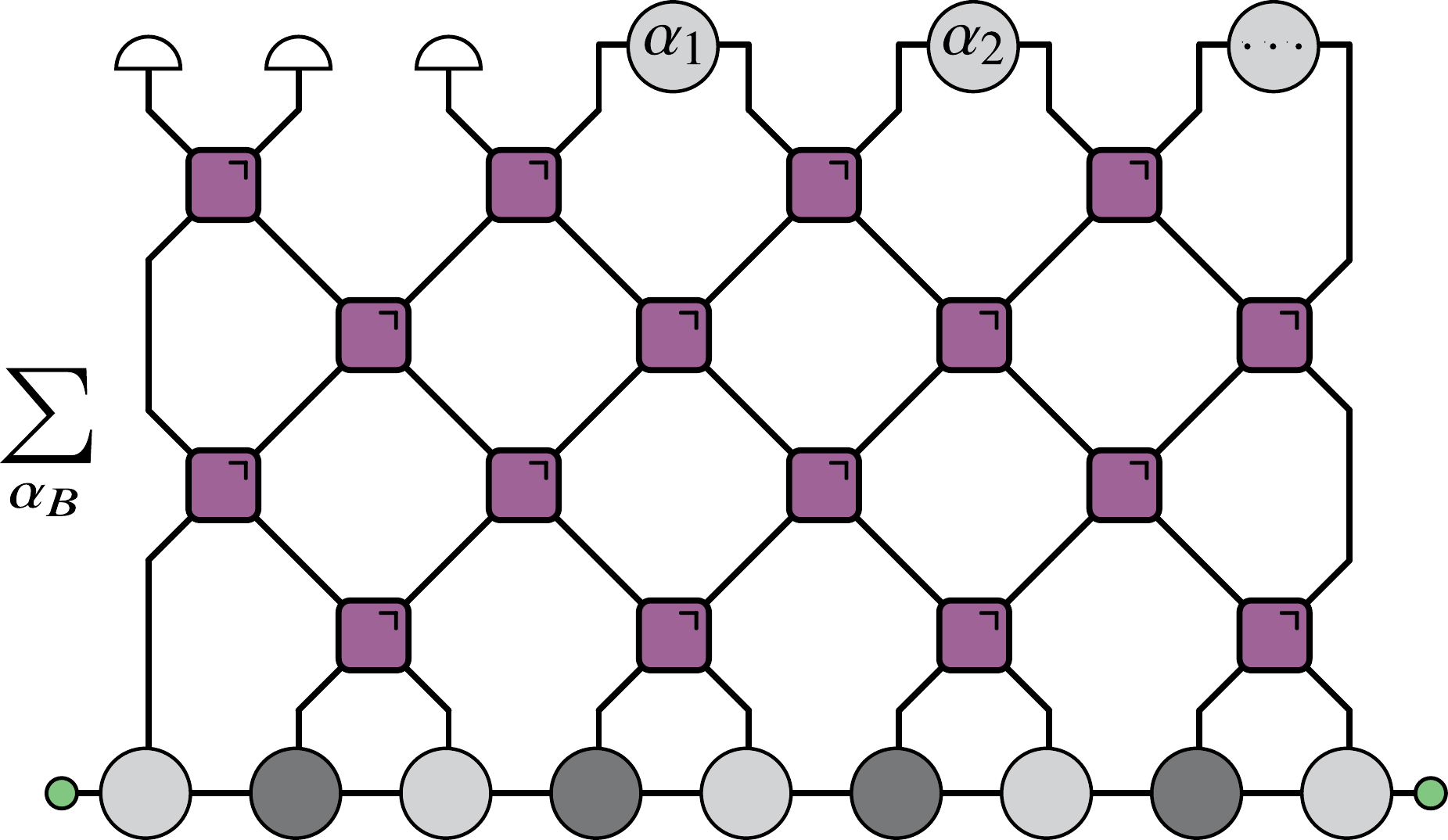}
\end{align}
where $\alpha_B = \{\alpha_1,\alpha_2, \dots, \alpha_{N_B/2}\}$. Here we have also introduced arbitrary boundary conditions on the MPS because we are working with closed boundary conditions, but the precise form of this boundary will be unimportant for what follows. It is important to note that all elements in this diagram correspond to folded versions of the previously introduced elements.

A transfer matrix can again be identified as
\begin{align}\label{eq:transfermat_MPS}
\includegraphics[width=0.7\linewidth]{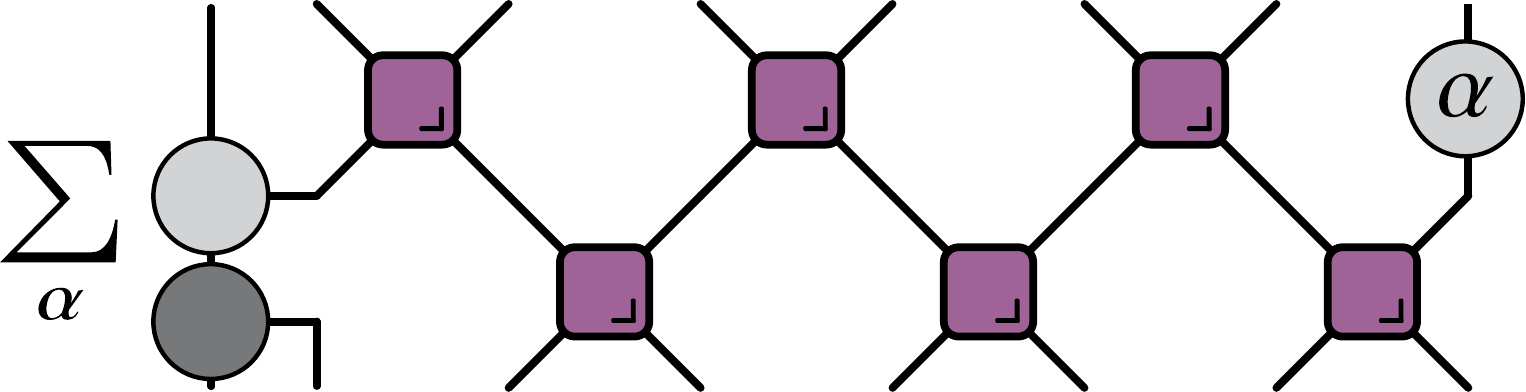}
\end{align}
and combining dual-unitarity and the solvability of both the initial state and the measurement basis leads to a set of eigenoperators
\begin{align}\label{eq:eigenstate_pi_MPS}
\vcenter{\hbox{\includegraphics[width=0.65\linewidth]{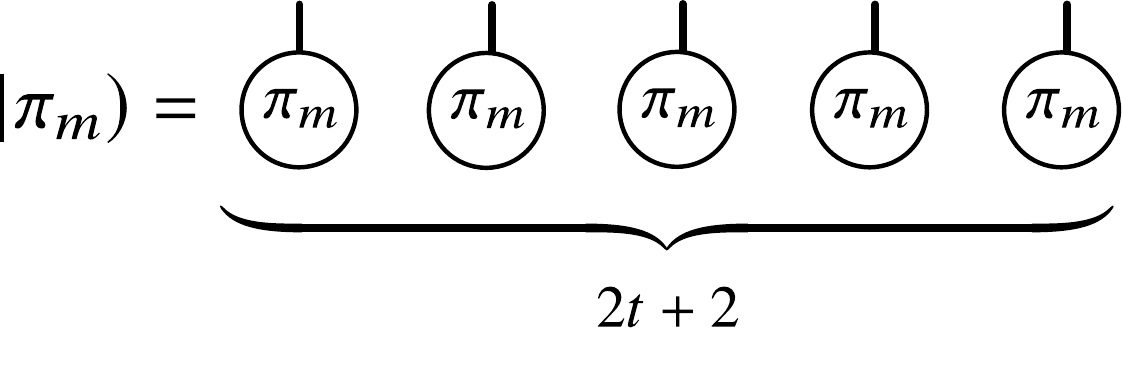}}}
\end{align}
again with eigenvalue $q^{2(1-m)}$. Note that the leftmost contractions act on the bond dimension Hilbert space, which is generally different from the physical Hilbert space, but this does not influence the following derivation. For $m=1$ this eigenstate can be illustrated by explicitly unfolding the diagram, leading to
\begin{align}
\vcenter{\hbox{\includegraphics[width=0.95\linewidth]{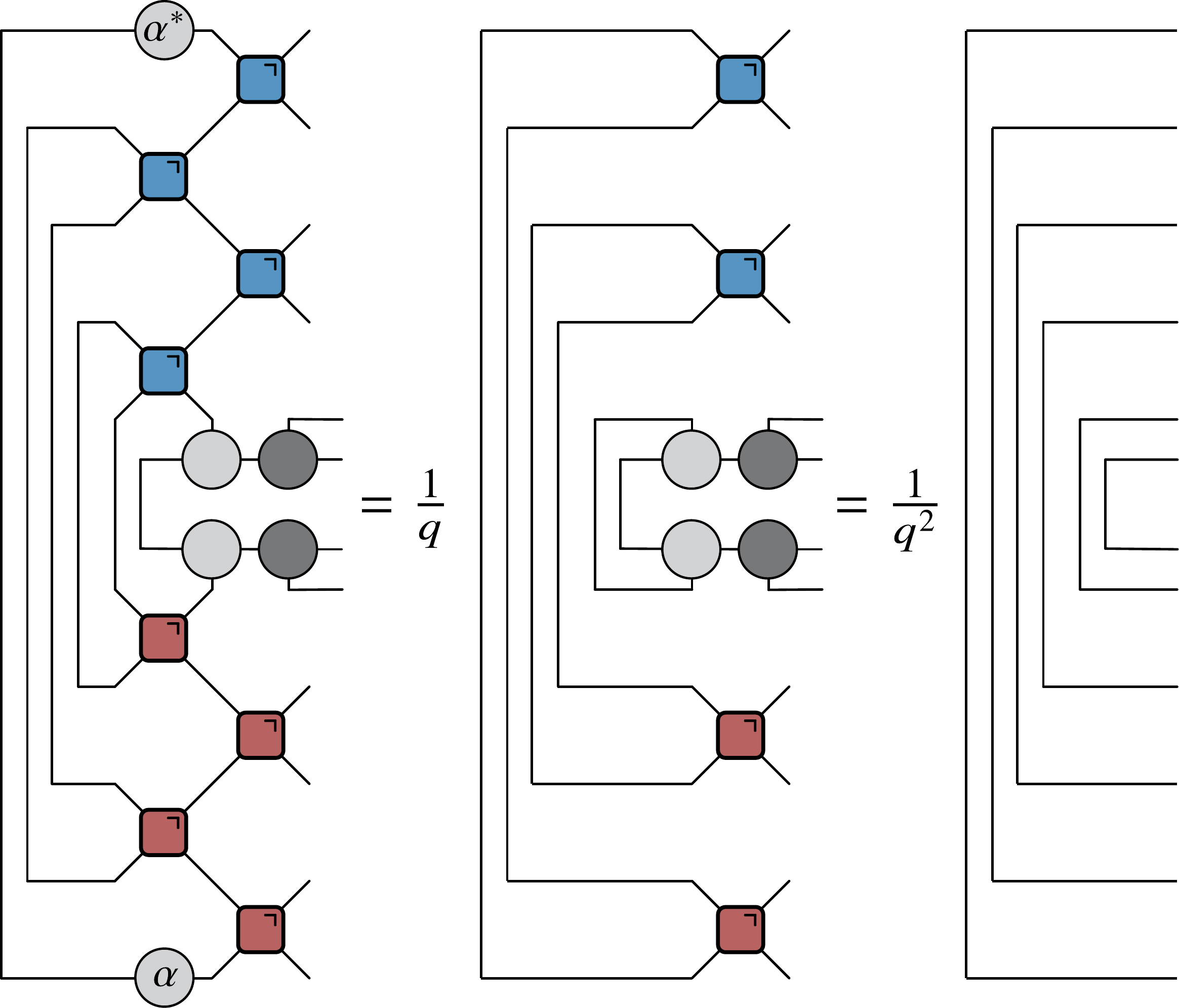}}}
\end{align}
In the first equality we have used the unitarity of $\alpha$ and the dual-unitarity of the underlying circuits, and in the second step the exact solvability condition \eqref{eq:solvabilityMPS} and the dual-unitarity. In the eigenvalue equation for the full transfer matrix the summation over two-site measurement outcomes $\alpha$ returns an additional prefactor $q^2$. 

These eigenoperators are both left and right eigenoperators of the full transfer matrix (as also observed for $m=1$ in Ref.~\cite{piroli_exact_2020}). We again assume that these permutation operators exhaust the leading eigenoperators for any $m$, since the appearance of additional eigenoperators would require additional structure not present in generic (e.g. maximally chaotic) dual-unitary circuits. The resulting derivation is identical to the one from Section~\ref{subsec:dualunitarity}, such that we can conclude that in these systems the projected ensemble again returns a quantum state design. The only difference is that the appearance of an additional contraction in the eigenoperators due the measurement outcome leads to an exact emergent state design at $t \geq N_A/2$ rather than $t > N_A/2$, as will also be demonstrated in the next section. From the above derivation we can conclude that measurements in a unitary error basis give rise to exact emergent quantum state designs for generic dual-unitary circuit dynamics starting from a solvable initial state (provided no additional leading eigenoperators exist, see next Section). 

Note that the emergent quantum state design from Sec.~\ref{subsec:Hadamard} can be directly related to this construction, since all gates satisfying Eq.~\eqref{eq:KIM_identity} map product states in the computational basis to maximally entangled states satisfying Eq.~\eqref{eq:solvablepairstate}. After a single application of a layer of complex Hadamard gates, the initial product state is mapped to a solvable state, and the one-site measurements in the computational basis are similarly mapped to solvable measurements.

A note about completeness: For $q=2$ all unitary error bases are equivalent to the Pauli matrices and hence the Bell pair states, but for higher dimensions of the local Hilbert space various constructions of unitary error bases have been proposed (see e.g. \cite{reutter_biunitary_2019}). This situation is similar to that for both dual-unitary gates and solvable initial states with small bond dimension: exhaustive parametrizations exist only for local qubits ($q=2$) \cite{bertini_exact_2019,piroli_exact_2020}. For larger dimensions of the local Hilbert space various constructions exist, either as direct generalizations of the qubit case \cite{claeys_ergodic_2021} or using more involved approaches \cite{borsi_remarks_2022}.

\section{Numerics}
\label{sec:numerics}
\begin{figure}[tb!]
\includegraphics[width=\columnwidth]{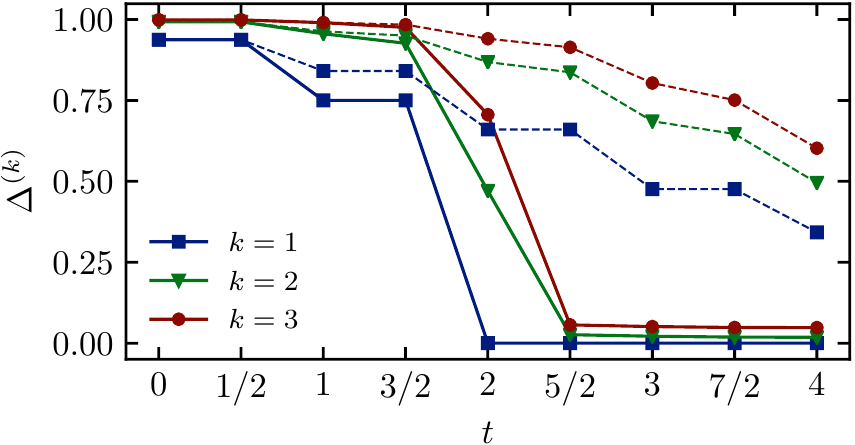}
\caption{Trace-norm distance between the projected ensemble and the Haar-random ensemble $\Delta^{(k)}$ for a system with $N_A=4$, $N_B=16$ for a number of time steps $t$. Full lines denote the results for the dual-unitary Hadamard circuit, dashed lines denote the results for a unitary circuit. Both systems are prepared and measured in the computational basis. \label{fig:numerics_zz}}
\end{figure}

\begin{figure}[tb!]
\includegraphics[width=\columnwidth]{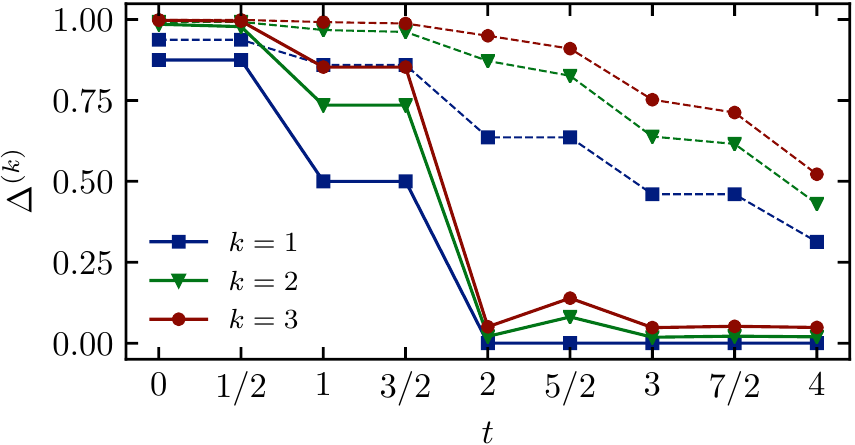}
\caption{Trace-norm distance between the projected ensemble and the Haar-random ensemble $\Delta^{(k)}$ for a system with $N_A=4$, $N_B=16$ for a number of time steps $t$. Both systems are evolved using identical (chaotic) dual-unitary circuit dynamics. Full lines denote the results for a system prepared in a solvable state and measured in a solvable basis, dashed lines for a system prepared and measured in a product state in the computational basis. \label{fig:numerics_BellvsZ}}
\end{figure}

In this section we numerically verify these constructions by explicitly comparing the projected ensemble to the Haar-random uniform ensemble, contrasting the solvable measurement schemes with generic ones. Numerically, the wave function is stored as a matrix product state, where unitary gate dynamics can be implemented using a TEBD algorithm. For the relatively short times considered here the time evolution can be performed exactly for a modest bond dimension, and for the finite bath size of Figs.~\ref{fig:numerics_zz} and \ref{fig:numerics_BellvsZ} the averaging over measurement outcomes can be performed exactly. All MPS calculations were performed using TeNPy \cite{hauschild_efficient_2018}.

The emergent quantum state design for complex Hadamard gates is illustrated in Fig.~\ref{fig:numerics_zz} for a system with $N_A=4$ and a limited bath size of $N_B=16$. The system is initially prepared in an eigenstate of the computational basis $\ket{00\dots0}$, is then evolved using a unitary circuit composed of either identical complex Hadamard gates or identical unitary gates, realizing Floquet dynamics in both cases. For $q=2$ the Hadamard circuits are equivalent to the KIM, where we fix $h_1=h_2=0.5$, away from the integrable point. The bath is measured using single-site measurements in the computational basis. In order to quantify the randomness, we consider the trace-norm distance between the resulting projected ensemble and the Haar-random ensemble (as was also done in Refs.~\cite{cotler_emergent_2021,ho_exact_2021}),
\begin{align}
\Delta^{(k)} = \frac{1}{2}||\rho_{\mathcal{E}}^{(k)}-\rho^{(k)}_{\textrm{Haar}}||\,.
\end{align}
For the Hadamard circuit all moments reach their steady-state value close to the Haar-random moments as soon as $t>N_A/2$, with a small but nonzero $\Delta^{(k)}$ because of the finite bath size. Increasing the bath size further reduces this final $\Delta^{(k)}$ value. For the unitary circuit no such collapse is observed and longer times are required to reach an emergent quantum $k$-design, with higher values of $k$ requiring longer times. Note that for the dual-unitary circuits considered here $\rho^{(1)}_{\mathcal{E}}=\rho_A$ exactly equals the infinite-temperature density matrix $\rho_A = \mathbbm{1}/q$ for $t \geq N_A/2$, irrespective of the measurement scheme, following the arguments of Ref.~\cite{gopalakrishnan_unitary_2019}. 

We also note the monotonicity of the trace distance to the projected ensemble: $\Delta^{(k)} \geq \Delta^{(k-1)}$. The $(k-1)$-th moments can be obtained by tracing out one copy of the Hilbert space in the $k$-th moments, which is a completely positive trace-preserving map. The trace distance is contractive under such maps, leading to the observed monotonicity\footnote{We thank the anonymous referee for this argument.}.

In Fig.~\ref{fig:numerics_BellvsZ} we compare the dynamics for two systems evolved using the same dual-unitary circuit, not equivalent to the KIM (the parametrization is given in Appendix \ref{app:num_details} for completeness). Rather than changing the circuit dynamics, we now consider systems that are either prepared and measured in the computational basis, or prepared and measured in the Bell pair basis. We choose $\mathcal{N}^{{(i,j)}} = \delta_{i,j}/\sqrt{q}$, leading to $\mathcal{W}(\mathcal{N}) = \mathbbm{1}$ and an initial state
\begin{align}
\ket{\psi_0} = \bigotimes_{i}\left(\frac{\ket{00}_{2i-1,2i}+\ket{11}_{2i-1,2i}}{\sqrt{2}}\right)\,.
\end{align}
Since we are dealing with an odd number of sites, we fix site $L$ to again be an eigenstate in the computational basis $\ket{0}_L$, but any state would return the same result. Performing the calculation, we again observe an emergent quantum state design only in the solvable basis, this time at $t=N_A/2$, where the computational basis results in nonzero $\Delta^{(k)}$. This result highlights how it is not just the dual-unitarity that underlies the emergent quantum state design, but rather the combination of the dual-unitarity with a solvable initial state and measurement scheme.

\section{Discussion}
\label{sec:discussion}

In this section we briefly discuss the requirement for a solvable measurement scheme. In all presented constructions, a transfer matrix along the spatial direction can be identified for the time-evolved state. For the class of complex Hadamard gates the transfer matrix for measurement outcomes $(z_0,z_1)$ reads
\begin{align}\label{eq:T_z0z1}
\vcenter{\hbox{\includegraphics[width=0.65\linewidth]{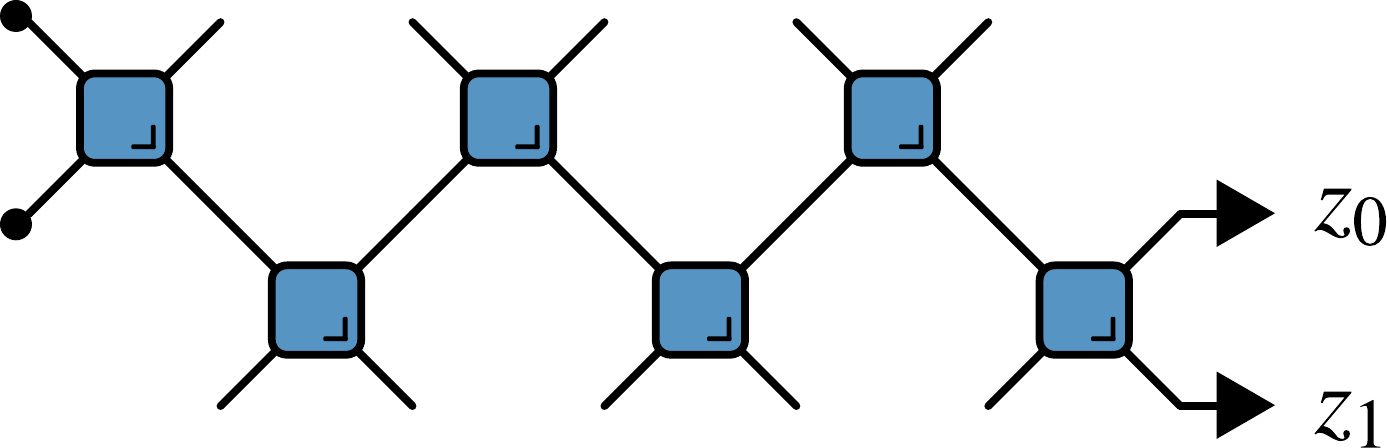}}}\,,
\end{align}
whereas the transfer matrix for a solvable MPS measurement scheme can be identified as
\begin{align}\label{eq:T_alpha}
\vcenter{\hbox{\includegraphics[width=0.65\linewidth]{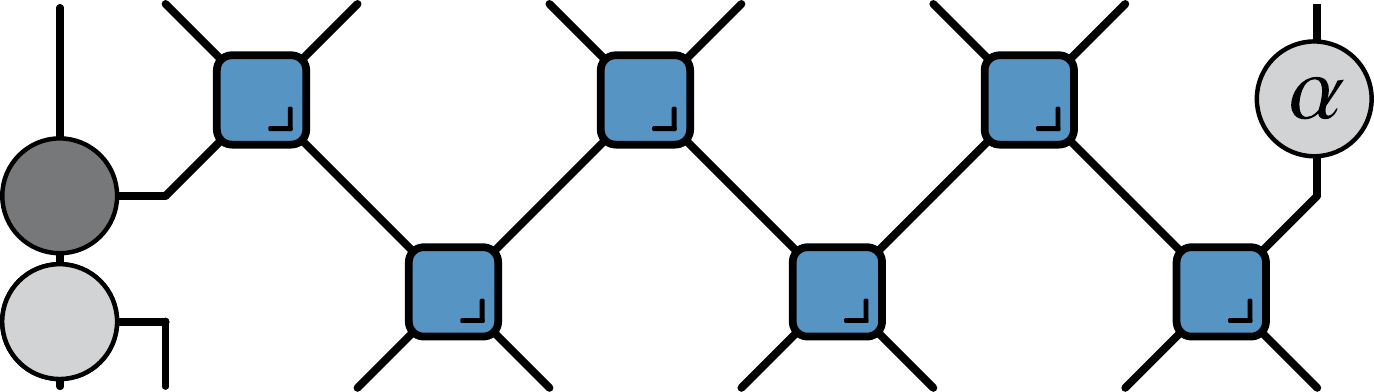}}}\,,
\end{align}
now depending on the measurement outcome $\alpha$. The measurement schemes in this work are designed such that for any measurement outcome these transfer matrices are (proportional to) unitary matrices, as a way of incorporating measurements without destroying the space-time duality of dual-unitary circuits. The unitarity of these transfer matrices guarantees that the folded transfer matrices [Eqs.~\eqref{eq:transfermatrix} and \eqref{eq:transfermat_MPS}] have a set of leading eigenstates $|\pi_m)$, since every contraction introduced by $|\pi_m)$ contracts a unitary operator with its hermitian conjugate.

However, a solvable measurement scheme also requires that these exhaust the leading eigenspace, i.e. no other eigenoperators with leading eigenvalue exist. Following the arguments of Ref.~\cite{ho_exact_2021}, this statement is equivalent to saying that the set of unitary transfer matrices for different measurement outcomes form a universal gate set, as proven for the specific case of the KIM away from certain fine-tuned points \cite{ho_exact_2021}. If this is the case, sampling the unitary transfer matrices over measurement outcomes is equivalent to sampling over Haar-uniform random unitary matrices, resulting in a quantum state design.

It is not guaranteed that the eigenoperators $|\pi_m) $ exhaust the leading eigenspace, and a proof would require the introduction of specific parametrizations for the dual-unitary gates and falls outside the aim of this work. We only note that the appearance of additional leading eigenoperators implies the existence of additional conservation laws for spatial evolution, precluding an exact quantum state design. Such additional eigenoperators naturally arise for nonergodic circuits exhibiting conservation laws, such as integrable dual-unitary circuits. 

While we cannot exclude the appearance of additional leading eigenoperators for generic dual-unitary circuits, i.e. without considering specific parametrizations, such eigenoperators seemingly require fine-tuning leading to additional structure, as also corroborated by numerical investigations. It was observed in Refs.~\cite{bertini_exact_2019,bertini_operator_2020} that generic dual-unitary circuits lead to chaotic dynamics, and we similarly expect emergent quantum state designs in generic dual-unitary dynamics under the presented measurement schemes.

\section{Conclusions}
\label{sec:conclusions}
Recent works have highlighted how projective measurements on a bath can give rise to an emergent quantum state design in unitary dynamics. Here we have established families of dual-unitary dynamics that are expected to give rise to an exact quantum state design in the thermodynamic limit of an infinite bath after a time proportional to the subsystem size. Such schemes require not just dual-unitarity, but also a solvable initial state and a solvable measurement scheme. 
Making use of the recent connection between dual-unitarity and biunitarity, we constructed two families of models where purely local measurements preserve the space-time duality of the circuit and are expected to lead to an exact quantum state design. The first introduces dual-unitary gates constructed out of complex Hadamard matrices as a direct extension of the kicked Ising model, incorporating previous parametrizations, for which product states in the computational basis act both as a solvable initial state and a solvable measurement. The second is based on the recent matrix product state construction of solvable initial states, and we show how unitary error bases can be used to construct solvable two-site measurements. Numerics were presented for qubit systems, but the presented constructions are applicable to arbitrary dimensions of the local Hilbert space. All presented results hold for both Floquet systems, where all gates are identical, and for circuits where all gates are different but chosen within the same class.

While we focused on emergent state designs, the presented constructions can be applied in other settings: for the complex Hadamard gates previous results on entanglement growth based on the KIM can be directly extended to these gates \cite{gopalakrishnan_unitary_2019}. The transfer matrices considered in this work are highly similar to the influence matrix determining temporal entanglement \cite{lerose_influence_2021,giudice_temporal_2021,lerose_overcoming_2022}, suggesting further studies of the effect of dual-unitarity in temporal entanglement. Furthermore, it is a natural question to wonder how measurements in a solvable basis differ from general measurements in the context of measurement-induced phase transitions \cite{li_quantum_2018,skinner_measurement-induced_2019,chan_unitary-projective_2019,gullans_dynamical_2020}, which have recently also been studied for dual-unitary circuits \cite{zabalo_operator_2022}

\textbf{Note added.} Shortly before this work was accepted for publication, Ref.~\cite{ippoliti_dynamical_2022} showed that for qubits and measurements in a Bell pair basis almost all dual-unitary gates lead to an exact emergent quantum state design.

\section*{Acknowledgments} 
We gratefully acknowledge support from EPSRC Grant No. EP/P034616/1. AL thanks Wen Wei Ho for explaining the results of Ref.~\cite{ho_exact_2021} during a stay at KITP, Santa Barbara. This research was supported in part by the National Science Foundation under Grant No. NSF PHY-1748958.

\appendix
\section{Dual-unitary gates from biunitarity}
\label{app:biunitary}
In this Appendix we explicitly relate the dual-unitarity of the Hadamard gate to the biunitarity of complex Hadamard matrices. This connection relies on the results from Ref.~\cite{reutter_biunitary_2019}, which presented a general overview of biunitary constructions in quantum information. Even though dual-unitarity was not considered, all results directly extend to dual-unitary gates. While biunitary two-site unitary circuits are dual-unitary, biunitarity is defined for a larger class of objects. In order to accommodate these objects, the graphical calculus from the main text needs to be extended by the inclusion of shaded regions. Any shaded region corresponds to an additional index, and all objects bordering this shaded region are additionally labelled by this same index. Indices corresponding to closed regions are implicitly summed over. After evaluation of the indices corresponding to shaded regions and removal of the shaded regions, the resulting diagrams can be evaluated in the same way as the diagrams from the main text. All graphical equalities should also be interpreted as equalities up to factors of $q$.

Biunitary models return the identity under both horizontal and vertical compositions, where the interpretation of these diagrams depend on the shaded regions (we refer the reader to Ref.~\cite{reutter_biunitary_2019} for a detailed introduction and discussion of non-dual-unitary biunitary constructions). A dual-unitary gate is represented in the usual way as
\begin{align}
\includegraphics[width=0.25\linewidth]{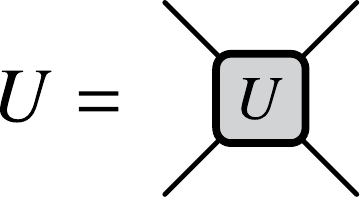}
\end{align}
where vertical and horizonal composition are respectively given by
\begin{align}
\includegraphics[width=0.4\linewidth]{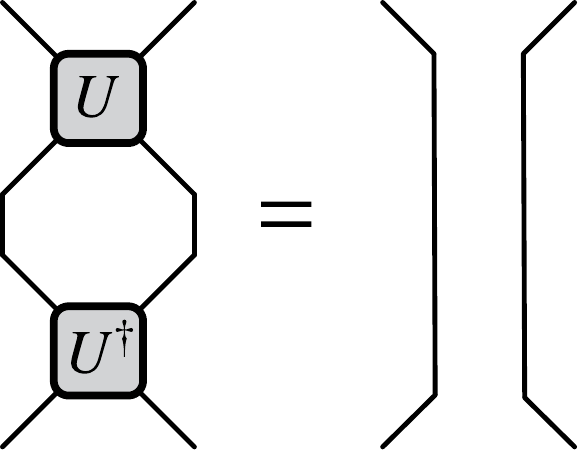}\qquad \includegraphics[width=0.35\linewidth]{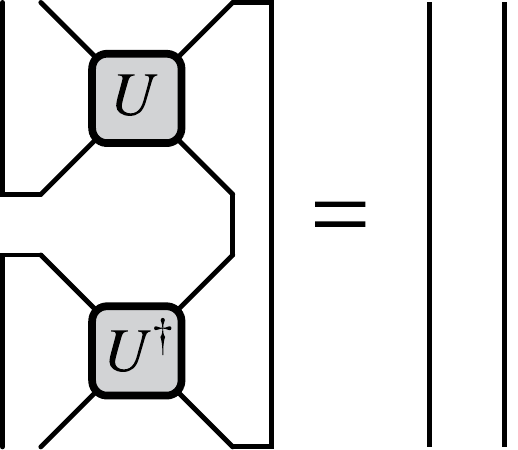}
\end{align}
Unitary error bases (UEBs) form a different biunitary construction, where we have a family of $q^2$ unitary matrices $U_n$ acting on a single copy of the local Hilbert space. These are graphically represented as
\begin{align}
\includegraphics[width=0.3\linewidth]{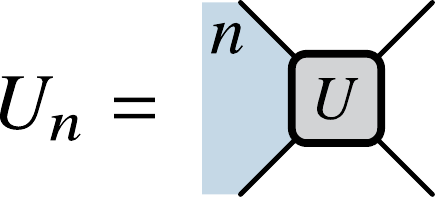}
\end{align}
Vertical composition here results in
\begin{align}
\includegraphics[width=0.4\linewidth]{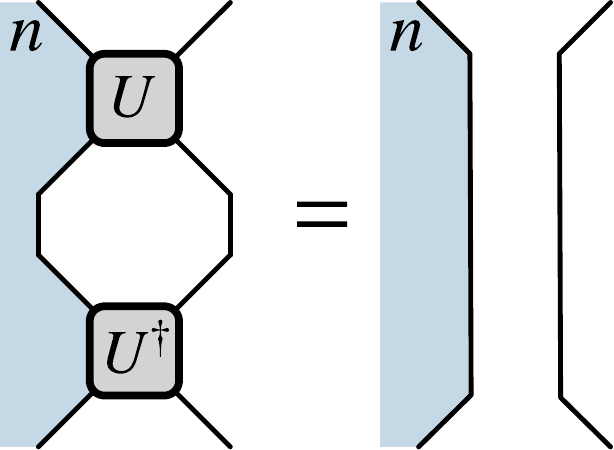},
\end{align}
representing the unitarity of the individual matrices, $U_n^{\dagger}U_n = \mathbbm{1}, \forall n$. Because of the asymmetry in the horizontal direction, horizontal composition leads to two different constraints,
\begin{align}
\includegraphics[width=0.35\linewidth]{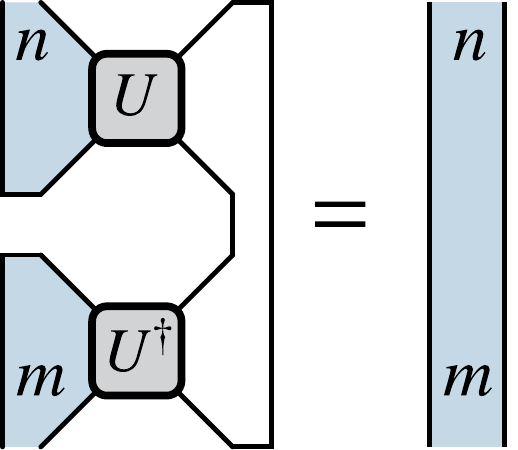}\qquad \includegraphics[width=0.35\linewidth]{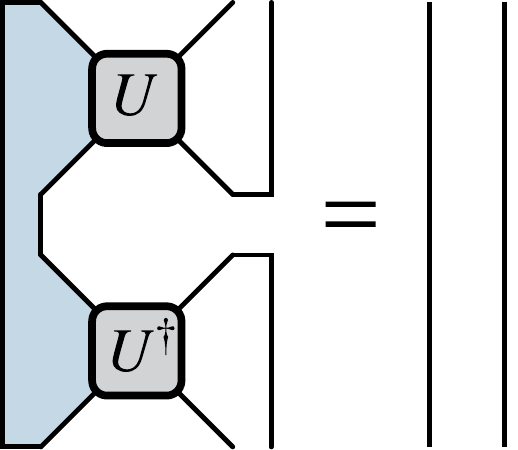}
\end{align}
In equation form, these represent the orthonormality and the completeness of the basis respectively,
\begin{align}
&\tr(U_n U_m^{\dagger}) = q \delta_{mn}\qquad\forall{m,n}, \nonumber\\
& \sum_{n} \big(U_n\big)_{ab}\big(U_n^{\dagger}\big)_{cd}= q\delta_{ad}\delta_{bc}\,.
\end{align}
Including two shaded regions returns complex Hadamard matrices $H \in \mathbbm{C}^{q\times q}$,
\begin{align}
\includegraphics[width=0.3\linewidth]{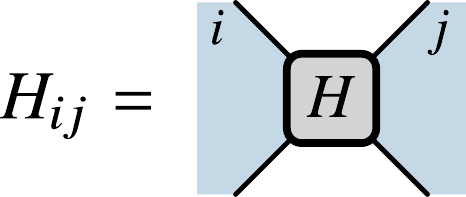}
\end{align}
The vertical and horizontal biunitary identities are given by
\begin{align}
\includegraphics[width=0.4\linewidth]{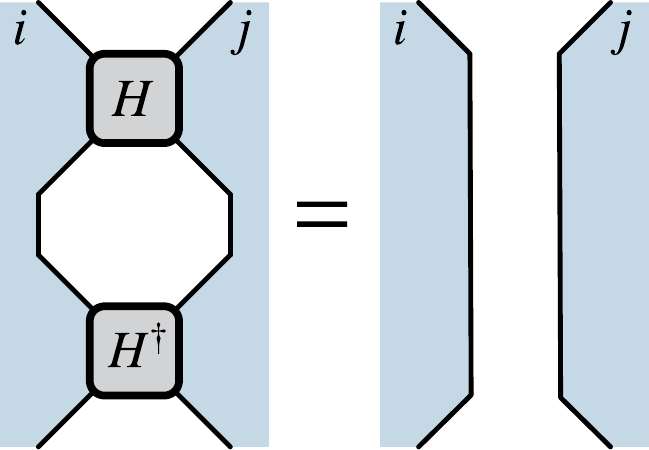}\qquad \includegraphics[width=0.35\linewidth]{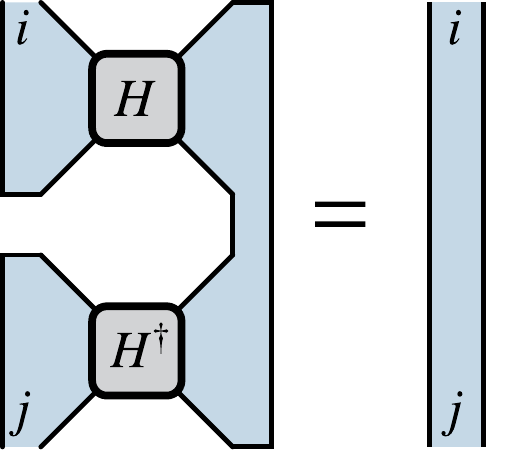}
\end{align}
leading to
\begin{align}
H_{ij}H^*_{ij} = 1, \, \forall i,j, \qquad \sum_k H_{ik}H^*_{jk} = q \delta_{ij}\,.
\end{align}
The graphical calculus makes clear in what way these objects satisfy similar properties. Quantum Latin squares can also be incorporated in this classification, but will not be considered in what follows. Ref.~\cite{reutter_biunitary_2019} showed how different biunitary connections can be combined to form new ones, which can here be used to construct dual-unitary gates from either complex Hadamard matrices or unitary error bases. A similar procedure to construct dual-unitary gates out of dual-unitary gates with a smaller dimension was recently discussed in Ref.~\cite{borsi_remarks_2022}.

First, we can consider the Hadamard gate from the main text: a dual-unitary gate acting on two sites with local Hilbert space $q$ constructed out of four $q\times q$ complex Hadamard matrices. Graphically, this gate is given by
\begin{align}\label{eq:constructionHad}
\includegraphics[width=0.65\linewidth]{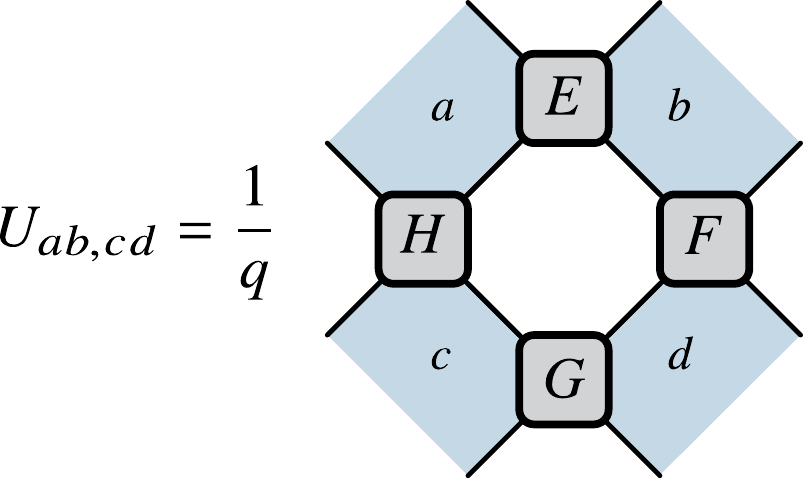}
\end{align}
The unitarity and dual-unitarity follow directly from the biunitarity of the building blocks, but can also be checked from the matrix elements $U_{ab,cd} = E_{ab}F_{bd}G_{dc}H_{ca}/q$. E.g. for unitarity we find
\begin{align}
&(U^{\dagger}U)_{ab,cd} = \sum_{e,f=1}^q U^*_{ef,ab}U_{ef,cd}  \nonumber\\
&= \frac{1}{q^2}\sum_{e,f=1}^q E^*_{ef}F^*_{fb}G^*_{ba}H^*_{ae} E_{ef}F_{fd}G_{dc}H_{ce} \nonumber\\
&= \frac{G^*_{ba}G_{dc}}{q^2}\left(\sum_{e=1}^qH^*_{ae}H_{ce}\right)\left(\sum_{f=1}^q F^*_{fb}F_{fd}\right)\nonumber\\
&= \delta_{ac}\delta_{bd}\,,
\end{align}
where we have used $|E_{ef}|^2=|G_{ba}|^2=1$ and the unitarity of the complex Hadamard matrices.

A highly similar parametrization was proposed in Ref.~\cite{gutkin_exact_2020}, which originally identified the connection between dual-unitary gates and Hadamard matrices. A class of gates was parametrized by two complex Hadamard matrices $F_{1,2}$, as
\begin{align}
U_{ab,cd} = \left(F_1\right)_{ab}\left(F_1\right)_{cd}\left(F_2\right)_{ac}\left(F_2\right)_{bd}/q\,.
\end{align}
It was also shown how these gates appear as the circuit representation of kicked dynamics, similar to the KIM. This expression can be interpreted as a specific case of Eq.~\eqref{eq:constructionHad} with $E=G=F_1$ and $F=H=F_2$. A specific case of a complex Hadamard matrix is the Fourier matrix, defined as
\begin{align}
K_{ab} = \exp\left[2\pi i a b / q\right]\,.
\end{align}
For $q=2,3,5$ all complex Hadamard matrices are equivalent to the Fourier matrix. Gutkin \emph{et al.} further studied the dynamics for gates where all complex Hadamard matrices equaled the Fourier matrix, $F_1=F_2=K$. A closely related `cat map' gate was proposed in Ref.~\cite{aravinda_dual-unitary_2021}, for which
\begin{align}
U_{ab,cd} = \frac{1}{q}&\exp\left[\frac{2\pi i}{q}\left(ab+cd+ac-bd\right)\right]\,.
\end{align}
This expression can again be written in the form of Eq.~\eqref{eq:constructionHad}, but now with $E=G=H=K$ and $F=K^*$. Our construction can also be contrasted with the results of Ref.~\cite{borsi_remarks_2022}, which already presented a generalization of the above examples constructed out of Fourier matrices by proposing a class of more general dual-unitary gates constructed out of Fourier matrices using `graph states', whereas our construction does not depend on the explicit parametrization of the complex Hadamard matrices.

All dual-unitary circuits constructed in this way satisfy the KIM property \eqref{eq:KIM_identity}, as can again be checked by direct calculation and making use of the properties of Hadamard matrices.

For completeness, we also note another construction based on biunitarity. A class of dual-unitary gates acting on a $q^2$-dimensional local Hilbert space can be found from 4 unitary error bases as
\begin{align}
\includegraphics[width=0.65\linewidth]{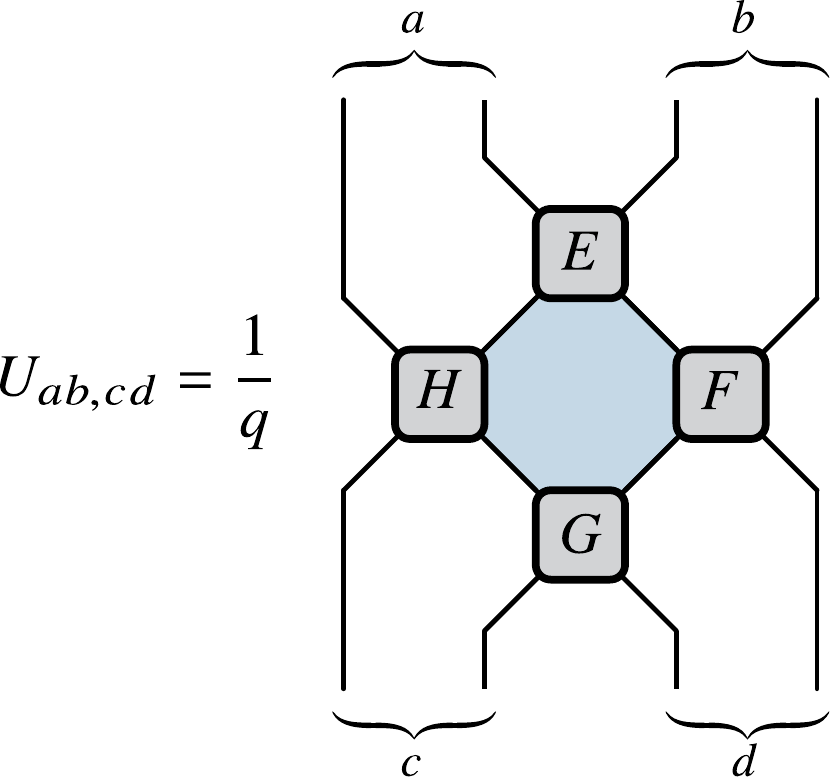}
\end{align}
The local $q^2$-dimensional Hilbert space now consists of a doubled $q$-dimensional Hilbert space for the UEB, which we can label as e.g. $a=(a_1,a_2)$. In matrix elements, this returns
\begin{align}
U_{ab,cd} = \frac{1}{q}\sum_{n=1}^{q^2}\left(E_n\right)_{a_2b_1}\left(F_n\right)_{b_2d_2}\left(G_n\right)_{d_1c_2}\left(H_n\right)_{c_1a_1}\,.
\end{align}
This construction is again explicitly symmetric when constructing the dual of $U$, such that unitarity implies dual-unitarity. While both are implicit in the graphical construction, they can also be explicitly verified from the matrix elements. For unitarity we have
\begin{widetext}
\begin{align}
\sum_{e,f=1}^{q^2} U^*_{ef,ab}U_{ef,cd}&=\frac{1}{q^2}\sum_{e_1,e_2,f_1,f_2=1}^q \sum_{m,n=1}^{q^2} \left(E_n^*\right)_{e_2f_1}\left(F_n^*\right)_{f_2b_2}\left(G_n^*\right)_{b_1a_2}\left(H_n^*\right)_{a_1 e_1} \nonumber\\
&\qquad \qquad \qquad \qquad \qquad   \times \left(E_m\right)_{e_2f_1}\left(F_m\right)_{f_2d_2}\left(G_m\right)_{d_1c_2}\left(H_m\right)_{c_1e_1} \nonumber \\
&=\frac{1}{q}\sum_{n=1}^{q^2}\left(\sum_{f_2=1}^q\left(F_n^*\right)_{f_2b_2}\left(F_n\right)_{f_2d_2}\right)\left(\sum_{e_1=1}^q \left(H_n^*\right)_{a_1 e_1} \left(H_n\right)_{c_1e_1}\right)\left(G_n^*\right)_{b_1a_2}\left(G_n\right)_{d_1c_2}  \nonumber\\
&=\delta_{a_1,c_1}\delta_{b_2,d_2}\times\frac{1}{q} \sum_{n=1}^{q^2}\left(G_n^*\right)_{b_1a_2} \left(G_n\right)_{d_1c_2}=\delta_{a_1,c_1}\delta_{a_2,c_2}\delta_{b_1,d_1}\delta_{b_2,d_2} = \delta_{ac}\delta_{bd}\,.
\end{align}
Here we have used that the summation over $e_{2},f_1$ returns $\tr(E_n^{\dagger}E_m) = q\delta_{mn}$ in the first equality, that $F_n$ and $H_n$ are unitary in the second equality, and that $G_n$ forms a complete basis in the third equality.


\section{Evaluating the boundaries}
\label{app:boundaries}

In order to evaluate $\rho^{(n,k)}_{\mathcal{E}}$ we need to evaluate the overlap of the left and right boundaries with the eigenoperators of the transfer matrix. Following the notation of the main text, we write
\begin{align}
\includegraphics[width=0.6\linewidth]{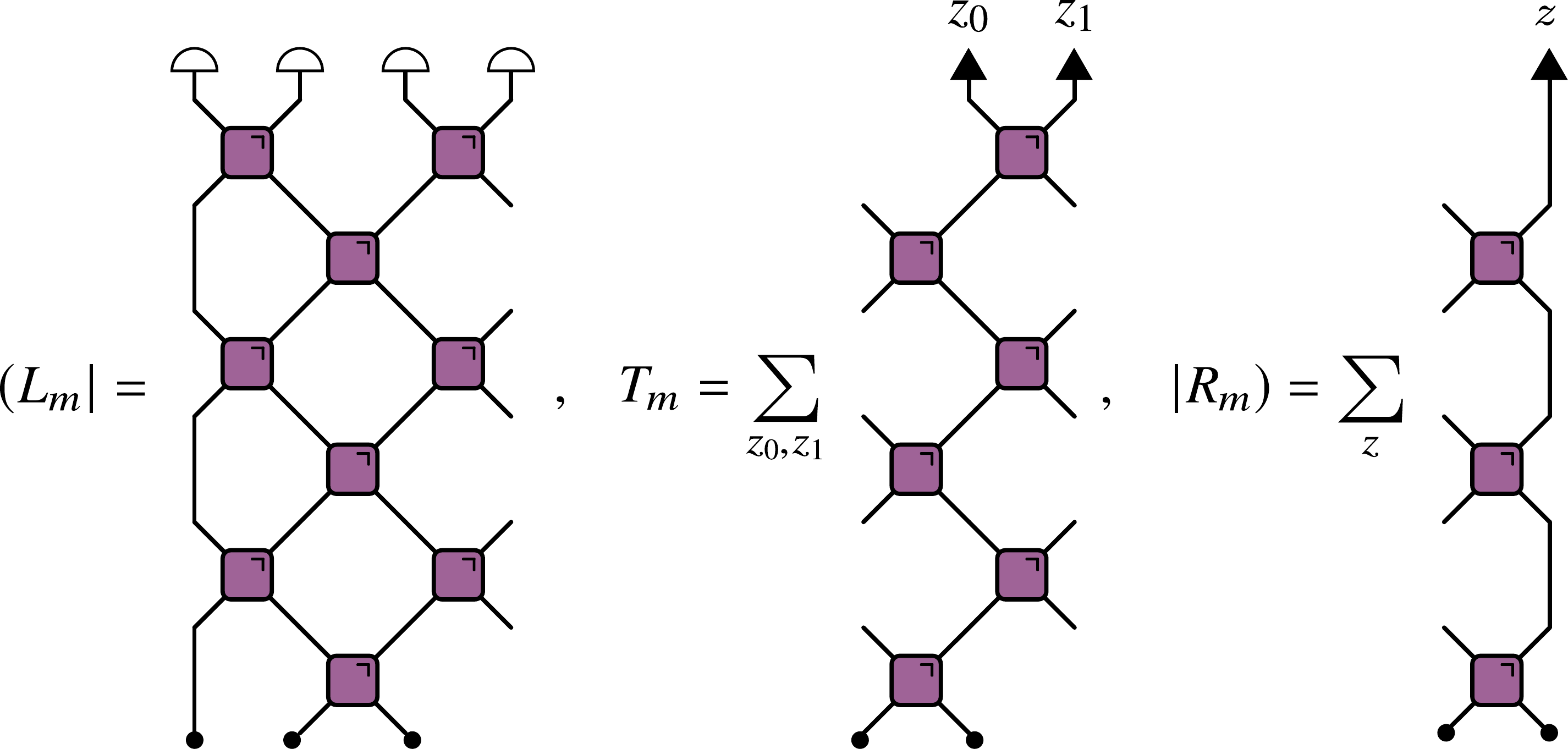}
\end{align}
\end{widetext}
For convenience, we denote the projector as
\begin{align}
\sum_{\sigma_m,\tau_m \in S_m}\textrm{Wg}(\sigma_m\tau_m^{-1},d_t) |\tau_m )(\sigma_m|,
\end{align}
with $d_t=q^{2t-1}$, and the left and right boundaries as $(L_m|$ and $|R_m)$ respectively. The overlaps $(\pi_m|R_m)$ can be directly evaluated by using dual-unitarity,
\begin{align}
\includegraphics[width=0.85\linewidth]{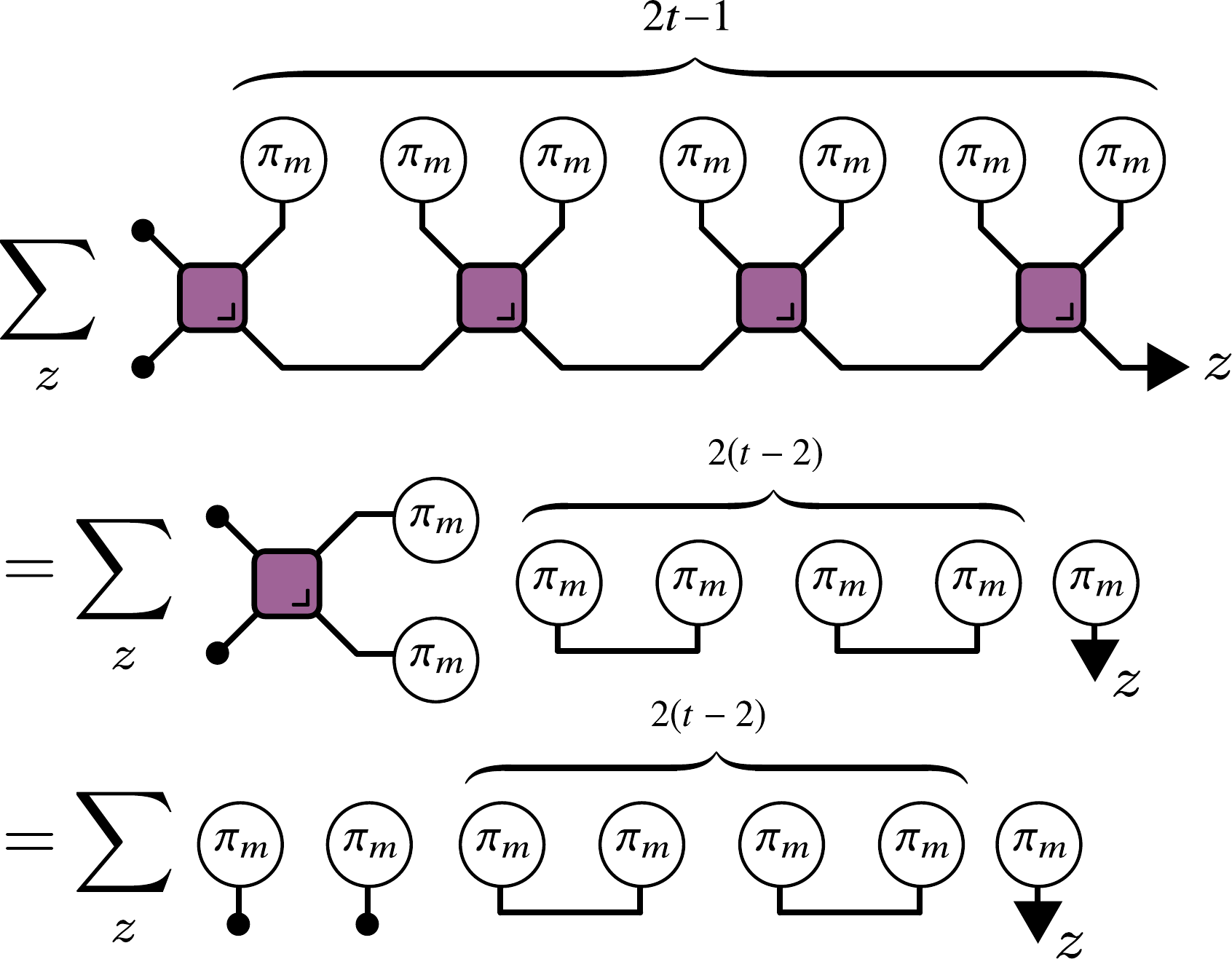}
\end{align}
to return an overlap of $(\pi_m|R_m)=q^{1+m(t-2)}$, independent of the permutation $\pi_m$. We find that
\begin{align}
&\sum_{\sigma_m,\tau_m \in S_m}\textrm{Wg}(\sigma_m\tau_m^{-1},d_t) (L_m|\tau_m )(\sigma_m|R_m) \nonumber\\
&\quad=\sum_{\sigma_m, \tau_m \in S_m}\textrm{Wg}(\sigma_m\tau_m^{-1},d_t)(L_m|\tau_m )\times  q^{1+m(t-2)} \nonumber\\
&\quad= \frac{q^{1+m(t-2)}}{d_t(d_t+1)\dots (d_t+m-1)}\sum_{\tau_m \in S_m}(L_m|\tau_m ) \,,
\end{align}
where we have used the property of the Weingarten functions
\begin{align}
\sum_{\sigma_m \in S_m}\textrm{Wg}(\sigma_m,d) = \frac{1}{d(d+1)\dots (d+m-1)}\,,
\end{align}
to evaluate the summation over $\sigma_m$.

For the left boundary, if $t \geq N_A/2$, we illustrate the evaluation of the overlap $(L_m|\pi_m)$ for $N_A=4$ and $t=3$ as
\begin{widetext}
\begin{align}
\includegraphics[width=0.95\linewidth]{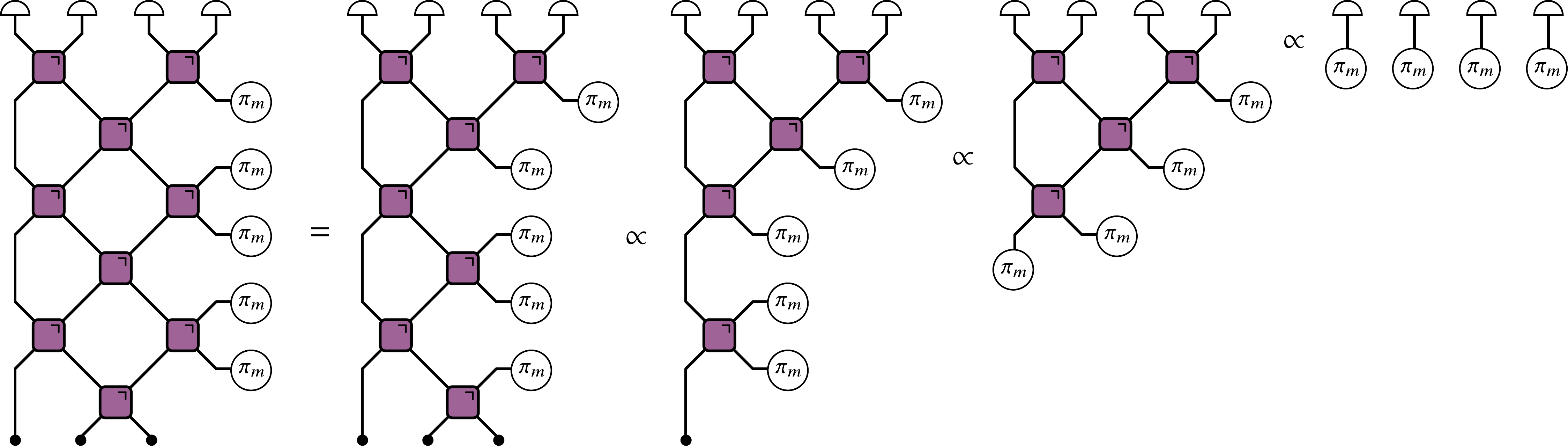}
\end{align}
\end{widetext}
In the first equality we have used the dual-unitarity, in the second Eq.~\eqref{eq:KIM_identity_layered}, and the remaining equalities only depend on the (dual-)unitarity. Such a simplification can always be performed for $t > N_A/2$, whereas if $ t\leq N_A/2$ not all gates can be removed using such contractions, and we obtain a nontrivial operator (this directly follows from the fact that there are less than $N_A$ contractions $\pi_m$ that can `flow' to the space direction). Note how the evaluation of these contractions requires the presence of a closed boundary, such that this approach fails for periodic boundary conditions.

For the final expression we need to include a summation over $\pi_m$. The resulting operator acts on $k$ copies of the subsystem Hilbert space, permuting different copies. Since this operator is explicitly symmetric under exchange of layers, it is necessarily proportional to the permutation operator acting on $k$ copies of the Hilbert space, i.e.
\begin{align}\label{eq:m_prop_k}
\includegraphics[width=0.95\linewidth]{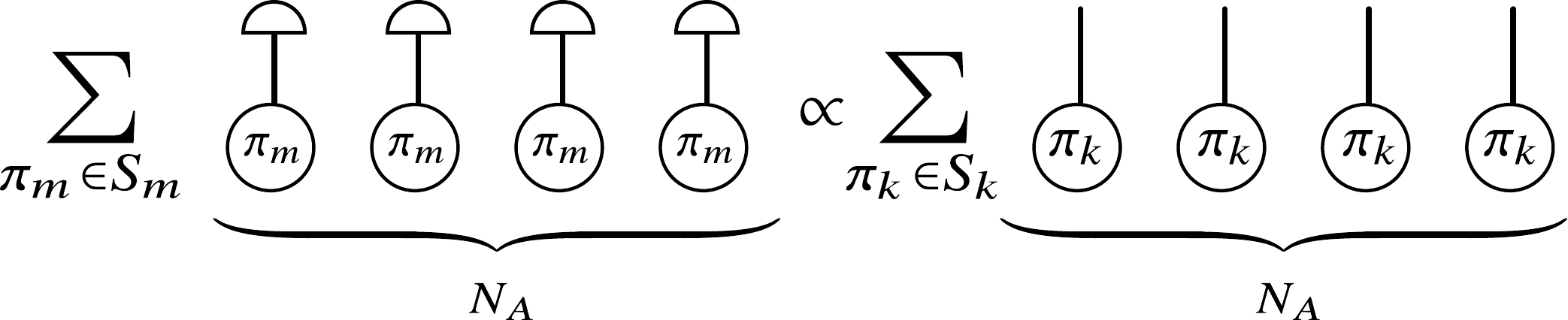}
\end{align}
The proportionality factor can be found by taking the overlap of both sides with the operator associated with the identity permutation on $k$ elements, returning the overlap with the identity permutation for $m$ elements on the left hand side. This overlap can be calculated using
\begin{align}
\includegraphics[width=0.35\linewidth]{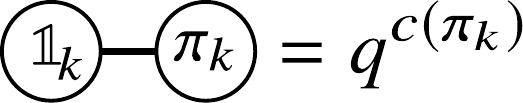}
\end{align}
where $c(\pi_k)$ is the number of cycles in $\pi_k$, as becomes apparent when unfolding this expression. The number of permutations of $m$ elements with $\ell$ cycles is given by the Stirling numbers of the first kind $\left[\begin{smallmatrix} m \\ \ell \end{smallmatrix}\right]$, which are the coefficients of the rising factorial
\begin{align}
d^{\overline{m}} = d (d+1) \dots (d+m-1) = \sum_{\ell=1}^{m}\left[\begin{smallmatrix} m \\ \ell \end{smallmatrix}\right]d^{\ell}\,.
\end{align}
For the right hand side of Eq.~\eqref{eq:m_prop_k}, we have that
\begin{align}
&\sum_{\pi_k \in S_k} q^{N_A \times c(\pi_k)} = \sum_{\ell=1}^{k}\left[\begin{smallmatrix} k \\ \ell \end{smallmatrix}\right]q^{N_A \ell} =d_A^{\overline{k}},
\end{align}
with $d_A = q^{N_A}$ the Hilbert space dimension for subsystem $A$. The left hand side similarly leads to $d_A^{\overline{m}} = d_A (d_A+1) \dots (d_A+m-1)$. Taking into account the additional powers of $q$ from the original contraction, we find that
\begin{align}
\sum_{\pi_m \in S_m}(L_m|\pi_m) = \frac{d_A^{\overline{m}}}{d_A^{\overline{k}}}\sum_{\pi_k \in S_k} P(\pi_k) \times q^{m(t-N_A)}\,.
\end{align}
Taking everything together returns the full projection as
\begin{align}
&\sum_{\sigma_m,\tau_m \in S_m}\textrm{Wg}(\sigma_m\tau_m^{-1},d_t) (L_m|\tau_m )(\sigma_m|R_m) \nonumber\\
&\qquad=\frac{d_A^{\overline{m}}}{d_t^{\overline{m}}d_A^{\overline{k}}}\sum_{\pi_k \in S_k} P(\pi_k) \times q^{m(2t-N_A-2)+1}\,.
\end{align}
This expression can now be directly evaluated at $m=1$ to return
\begin{align}
\frac{1}{d_A^{\overline{k}}}\sum_{\pi_k \in S_k} P(\pi_k)= \rho^{(k)}_{\textrm{Haar}}\,.
\end{align}

\vspace{0.5\baselineskip}
\section{Numerical details}
\label{app:num_details}
The dual-unitary gate used for Fig.~\ref{fig:numerics_BellvsZ} in Section~\ref{sec:numerics} is constructed using the approach outlined in Refs.~\cite{bertini_exact_2019,claeys_ergodic_2021} and this gate and the random unitary for Fig.~\ref{fig:numerics_zz} are given by
\begin{widetext}
\begin{align}
U_{\textrm{dual-unitary}} &= 
\begin{pmatrix}
0.5977+0.2173i & -0.0164+0.1243i & -0.6726-0.1356i & 0.2964+0.1454i \\
-0.5313-0.0565i &  0.3609-0.405i & -0.5851-0.1578i & 0.0152-0.2295i \\
0.0767+0.2142i & -0.2352-0.0701i &  0.1878-0.1941i & 0.4311-0.7932i\\
-0.0311-0.5072i & -0.7264-0.3196i & -0.1361-0.2654i & 0.0248+0.1498i
\end{pmatrix}	\\
U_{\textrm{unitary}} &= 
\begin{pmatrix}
-0.1284-0.1396i & 0.3885+0.2554i & -0.0213+0.0633i & -0.8157+0.2793i \\
-0.5712 -0.0761i & 0.6242-0.1894i & 0.1094-0.0217i & 0.1847-0.4426i \\
-0.2477-0.3292i & -0.3735-0.1889i & 0.6304 +0.4942i & -0.1150+0.0140i \\
0.4111 -0.5415i & 0.1245-0.4095i & 0.2321-0.5362i & -0.1046-0.0466i
\end{pmatrix}
\end{align}
\end{widetext}

\bibliographystyle{apsrev4-1}
\bibliography{Library.bib}

\end{document}